\documentclass[journal=jctcce,manuscript=article]{achemso}

\usepackage[version=3]{mhchem} 
\usepackage{braket}
\usepackage{mathtools}
\usepackage{amsthm, amssymb, amsfonts}
\usepackage{cancel}
\usepackage{array}
\usepackage{listings}
\usepackage{float}
\usepackage{bm}
\usepackage{graphicx}
\usepackage{threeparttable}
\usepackage[hidelinks]{hyperref}

\newcommand{\ketCC}{\left| \mathrm{CC} \right\rangle }

\author{Xiang Yuan}
\affiliation[PhLAM]{Univ. Lille, CNRS, UMR 8523 - PhLAM - Physique des Lasers Atomes et Molécules, F-59000 Lille, France}
\alsoaffiliation[Vrije Universiteit Amsterdam]
{Department of Chemistry and Pharmaceutical Sciences, Faculty of Science, Vrije Universiteit Amsterdam, 1081 HV Amsterdam, The Netherlands}

\author{Loïc Halbert}
\affiliation[PhLAM]{Univ. Lille, CNRS, UMR 8523 - PhLAM - Physique des Lasers Atomes et Molécules, F-59000 Lille, France}

\author{Johann Valentin Pototschnig}
\affiliation[Vrije Universiteit Amsterdam]
{Department of Chemistry and Pharmaceutical Sciences, Faculty of Science, Vrije Universiteit Amsterdam, 1081 HV Amsterdam, The Netherlands}

\author{Anastasios Papadopoulos}
\affiliation[Vrije Universiteit Amsterdam]
{Department of Chemistry and Pharmaceutical Sciences, Faculty of Science, Vrije Universiteit Amsterdam, 1081 HV Amsterdam, The Netherlands}

\author{Sonia Coriani}
\affiliation{DTU Chemistry -- Department of Chemistry, Technical University of Denmark, DK-2800 Kongens Lyngby, Denmark}

\author{Lucas Visscher}
\affiliation[Vrije Universiteit Amsterdam]
{Department of Chemistry and Pharmaceutical Sciences, Faculty of Science, Vrije Universiteit Amsterdam, 1081 HV Amsterdam, The Netherlands}

\author{André Severo Pereira Gomes}
\email{andre.gomes@univ-lille.fr}
\affiliation[PhLAM]{Univ. Lille, CNRS, UMR 8523 - PhLAM - Physique des Lasers Atomes et Molécules, F-59000 Lille, France}

\title[An \textsf{achemso} demo]
  {Formulation and Implementation of Frequency-Dependent Linear Response Properties with Relativistic Coupled 
  Cluster Theory for GPU-accelerated Computer Architectures}

\abbreviations{IR,NMR,UV}
\keywords{American Chemical Society, \LaTeX}

\begin{document}

\begin{abstract}
We present the development and implementation of the relativistic coupled cluster linear response theory (CC-LR)  
which allows the determination of
molecular properties arising from time-dependent or time-independent electric, magnetic, or mixed electric-magnetic perturbations (within a common gauge origin for the magnetic properties), as well as to take into account the finite lifetime of excited states in the framework of damped response theory. We showcase our implementation, which is capable to offload the computationally intensive tensor contractions characteristic of coupled cluster theory onto graphical processing units (GPUs), in the calculation of: \textit{(a)} frequency-(in)dependent dipole-dipole polarizabilities of IIB atoms and selected diatomic molecules, with a particular emphasis on the calculation of valence absorption cross-sections for the I$_2$ molecule;\textit{(b)} indirect spin-spin coupling constants for benchmark systems such as the hydrogen halides (HX, X = F-I) as well the H$_2$Se-H$_2$O dimer as a prototypical system containing hydrogen bonds; and 
\textit{(c)} optical rotations at the sodium D line for hydrogen peroxide analogues (H$_{2}$Y$_{2}$,  Y=O, S, Se, Te). Thanks to this implementation, we are able show the similarities in performance--but often the significant discrepancies--between CC-LR and approximate methods such as density functional theory (DFT). Comparing standard CC response theory with the flavor based upon the equation of motion formalism, we find that, for valence properties such as polarizabilities, the two frameworks yield very similar results across the periodic table as found elsewhere in the literature; for properties that probe the core region such as spin-spin couplings, on the other hand, we show a progressive differentiation between the two as relativistic effects become more important. Our results also suggest that as one goes down the periodic table it may become increasingly difficult to measure pure optical rotation at the sodium D line, due to the appearance of absorbing states.   
\end{abstract}

\section{Introduction}

The fundamental molecular properties, that are 
connected to the response of a system to external perturbations such as electric or magnetic fields, are central to the study of linear and non-linear optics \cite{barron2009molecular,bishop_molecular_1990,papadopoulos_non-linear_2006,cronstrand_multi-photon_2005}. It is widely acknowledged that molecules containing heavy elements, that is, those found towards the lower parts of the periodic table, have a plethora of applications. For instance, by manipulating the molecular polarizability, researchers can design materials with advanced optical properties for use in photovoltaic devices and glasses, such as bismuth oxide-based materials\cite{komatsu2020review}. Another important example is the utilization of optical activity to design Lanthanide complexes as chiral probes for biological processes\cite{carr2012lanthanide}. A detailed understanding of the physical phenomena behind these properties at the atomic or molecular level is very important to tune them or to provide insight for the development of new materials and novel applications.

In quantum mechanics, molecular properties can be derived via perturbation theory, or more specifically, through the response theory formalism, which 
in general lines 
identify molecular properties from the derivatives of the energy (or an equivalent quantity) with respect to the external perturbations. The genesis of modern response theory may perhaps be traced back to the introduction by~\citet{langhoff1972aspects} in 1972, of a formalism that allowed both time-dependent and time-independent perturbations to be taken into account analytically, 
i.e., without employing finite-difference (finite-field) approaches, which are numerically straightforward (but computationally expensive) and only applicable to the time-independent case. 
Among the properties one can calculate, those 
related to the linear response~\cite{norman_perspective_2011,helgaker_recent_2012} of the systems are particularly interesting since they give rise to e.g.\ the polarizability and optical activity, and can provide us with information on electronically excited states.

The current formulations of response properties may be categorized into those employing either Ehrenfest theorem~\cite{dalgaard1980time,olsen_linear_1985} 
or quasi-energy approaches\cite{rice1991calculation,sasagane1993higher,Christiansen1998}. Although response theory based on exact wave functions can provide the expressions for molecular properties directly, practical applications require the use of approximate models such as Hartree-Fock (HF) and density functional theory (DFT), and many other wave function based approaches such as multi-configuration self-consistent field (MCSCF), configuration interaction (CI), coupled cluster (CC) to name just a few (see~\citet{helgaker_recent_2012} for a comprehensive survey). To date response theory has achieved great success in dealing with a wide variety of molecular properties, and treating both small and large-scale systems \cite{pawlowski_molecular_2015,helgaker_recent_2012,norman_perspective_2011,norman_principles_2018,albota1998design,macak2000electronic}. Here, the availability of analytic derivatives approaches has proven to be important for efficient calculations, particularly for large-scale molecule simulations. 

However, while most formulations (and implementations) of response theory mentioned above are based on non-relativistic quantum mechanics, it is now widely recognized that when dealing with molecules containing heavy elements, relativistic effects must also be taken into account\cite{shee_analytic_2016,saue_relativistic_2011,vicha_relativistic_2020,bolvin_alternative_2006}. In addition, heavy elements also have more electrons than their lighter counterparts, which can bring about subtler effects due to electron correlation that may significantly impact the molecular properties. In the domain of relativistic quantum chemistry, the linear-response function based on approximate models including HF~\cite{saue_linear_2003,visscher_4-component_1997}, DFT~\cite{saue2002four,aquino_electric_2010}, and Second-Order-Polarization-Propagator Approximation (SOPPA)\cite{schnack-petersen_second-order-polarization-propagator-approximation_2020} has been well-established. Due to its modest computational cost, DFT has become the most widely used approach for correlated electronic structure theory, even though it is not possible to systematically improve the quality of calculations with currently available density functional approximations~\cite{burke2012perspective}. Due to that, depending on the property of interest, DFT results may deviate strongly from experimental or accurate theoretical models for relativistic electronic structure calculations, even for closed-shell species around the ground-state equilibrium structure~\cite{kervazo_accurate_2019,sunaga_towards_2021}. An alternative to DFT is found in CC theory, which is considered as a “gold standard”~\cite{crawford2007introduction,bartlett2007coupled} among electronic structure methods due to its ability to yield results that approach chemical accuracy for both correlation energies and properties. 

To date, there are various CC linear-response (CC-LR) implementations based on standard models such as CC2~\cite{Christiansen1995}, CCSD\cite{Christiansen1998,pawlowski_molecular_2015} and CC3~\cite{hald_calculation_2003}. These approaches have been shown to achieve good agreement with experimental values for both electric and magnetic molecular properties~\cite{crawford2019reduced,krylov2008equation,khani_uv_2019,gauss_coupledcluster_1995,christiansen_integral-direct_1998,ruud_optical_2002}. We also note the emergence in recent years of response theory implementations based on the equation-of-motion coupled cluster (EOM-CC) model~\cite{coriani_molecular_2016,nanda_communication_2018,alessio_equation--motion_2021,andersen2022cherry,andersen2022probing,faber2018resonant}, which are appealing due to their simpler programmable expressions while yielding exactly the same excitation energies as CC-LR, and nearly equivalent numerical results for response properties. In the time-dependent framework, as pointed out by~\citet{coriani_molecular_2016}, the EOM-CC response is equivalent to the combination of an exponential parametrization for the ground-state wavefunction, and a linear parametrization for the time-dependent wavefunction (which these authors refer to as a CC-CI type wave function), as opposed to the CC-LR case, which employs exponential parametrizations for both time-dependent and time-independent wavefunctions (referred to as CC-CC type wavefunctions). 

A significant downside of these implementations, however, is that they are available only for non-relativistic or rather approximate relativistic Hamiltonians. As such, they are not generally suitable for treating molecular systems containing heavy elements. In this manuscript, we aim to bridge this gap and present the implementation and pilot applications of CC-LR and EOM-CC models in combination with relativistic Hamiltonians, as part of the ExaCorr~\cite{pototschnig_implementation_2021} module of the DIRAC program~\cite{saue2020dirac}. One feature of ExaCorr is its ability, through the use of the ExaTENSOR\cite{lyakh_domainspecific_2019} library, to carry out distributed tensor operations with offloading to graphical processing units (GPUs)--which have been shown to be ideally suited to accelerate coupled cluster calculations due to the latter's substantial floating-point operation and memory-intensive nature~\cite{deprince2011coupled,calvin2020many,hohenstein2021gpu,pototschnig_implementation_2021,hillers2023massively}. In the work detailed here we take advantage of GPU offloading and thread-level parallelism, and will discuss the currently ongoing work to enable large-scale parallel calculations in a subsequent publication.

Apart from discussing our implementation, we showcase its generality and versatility by examining examples of three distinct classes of molecular properties: those involving purely electrical perturbations, purely magnetic perturbations, and mixed electric and magnetic perturbations. 

As an example of the first class, we take the electric dipole polarizabilities because of their significance in a wide range of applications and because they provide valuable insights into the properties and behavior of molecules. For example, materials with high dipole polarizabilities and dielectric constant are used in the polymers that are needed for high-energy-density capacitors\cite{thakur2016recent}, while materials with low dipole polarizabilities\cite{volksen2010low} are used as insulators in electrical devices. For optical spectroscopies, in the calculation of resonant processes such as electronic excitations it is important, from both a practical and physical points of view, to account for the finite excited-state lifetimes in the calculation of response functions, since these will relate to the broadening in the measured spectra. The damped coupled cluster response theory has in recent years emerged as a very effective tool for incorporating such effects in simulating the spectroscopy of complex molecules\cite{norman2001near,norman2005nonlinear,coriani2012coupled,coriani2012lanczos,kauczor2013communication}. In this manuscript, we demonstrate our ability to calculate damped response functions, as we can handle perturbing external fields with either real or complex frequencies.

We consider indirect nuclear spin-spin coupling constants as a representative of the second class. Indirect nuclear spin-spin coupling constants 
manifest themselves in Nuclear Magnetic Resonance (NMR) spectroscopy, which alongside optical spectroscopies is another invaluable tool in chemistry. As a substantial fraction of the atoms in the periodic table is NMR-active, the technique can very often be used to provide critical information about their chemical environment\cite{helgaker1999ab,helgaker2008quantum,vaara2007theory} in a non-destructive way. Regarding computational analysis, apart from the fact that theoretical calculations are extremely useful to interpret experimental signals, it has been demonstrated that it is essential to account for relativistic effects already for elements around the third row of the periodic table\cite{visscher1999full,franzke2021nmr,franzke2023reducing,aucar2018foundations,liu2017handbook}. Magnetic properties are often challenging to calculate, due to the dependence of the results on the gauge origin of an external magnetic field for incomplete bases sets. However, the indirect spin-spin coupling is expressed as the second derivative of the electronic energy with respect to the internal magnetic fields caused by nuclear spins, so that the gauge-origin issue does not arise.

Optical rotation is taken as an example of the third class. Studying optical rotation is of significant interest for several reasons. First and foremost, optical rotation measurements can provide information about the chiral nature of molecules. This is particularly important in the pharmaceutical industry, as many drugs are chiral and their properties can vary depending on their handedness\cite{nguyen2006chiral}. In addition to its applications in the pharmaceutical industry, studying optical rotations can also provide insights into the electronic and structural properties of molecules. Optical rotations are influenced by a variety of factors, including the electronic structure of the molecule, the molecular geometry, and the surrounding environment.  Moreover, in materials science, the optical properties of materials can be used to design and develop new materials\cite{he2008multiphoton}. For this property the gauge-origin issue mentioned above also arises\cite{helgaker_recent_2012}. In subsequent work we will explore approaches to ensure gauge-invariance for  coupled-cluster calculations of optical rotation~\cite{Pedersen2004,Caricato2020,Parsons2023}, but we note that for the small, symmetric molecules studied here the use of a common gauge origin yields sufficiently accurate results to allow for a comparison of different electronic structure approaches~\cite{Ruud2002,Ruud2003,Crawford2009}, which is our goal here.

The manuscript is organized as follows: In  
Section~\ref{Theory},
response theory and its corresponding parametrization for time-dependent coupled cluster wave-function are summarized. 
In Section~\ref{Implementation}, we described the details of the implementation. 
Section~\ref{ComputDetails} is devoted to the details of the computations we used to test the implementation. The calculations are presented and discussed in Section~\ref{Results}. Finally, a brief summary is given in Section~\ref{Conclusions}.

\section{Theory}
\label{Theory}

We base the theory on the 
time-averaged quasienergy formalism, which we briefly summarize below, and refer the reader to the landmark paper by~\citet{Christiansen1998} for a detailed discussion on it, as well as other more recent works~\cite{coriani_molecular_2016,pawlowski_molecular_2015,norman2018principles}.

\subsection{Response functions based on time-average quasienergy}

We aim to solve the time-dependent wave equation 
\begin{equation}
 i \frac{\partial}{\partial t} |\Psi(t)\rangle = H |\Psi(t) \rangle
\end{equation}
where ${H}$ is the total electronic Hamiltonian 
\begin{equation}
H = H_0 + V(t),
\end{equation}
composed of $H_0$, which represents the time-independent electronic Hamiltonian (e.g.\ the Dirac-Coulomb Hamiltonian, the eXact 2-component Hamiltonian (X2C), the Levy-Leblond Hamiltonian, etc., see~\cite{saue_relativistic_2011,saue2020dirac} and references therein), and $V(t)$ representing a sum of $N$ perturbations that are periodic in time with frequencies $\omega_k$
\begin{equation}
V(t) = \sum_{k=1}^N \left[ (e^{i\omega_k t} + e^{-i\omega_k t}) \sum_x \epsilon_x (\omega_k) {X} \right]
\end{equation}
expressed in terms of a one-body operator ${X}$ 
and the associated frequency-dependent perturbation strength $\epsilon_x (\omega_k)$. 
In the present study, ${X}$
corresponds, for instance, to the $x$-component of the electric dipole operator $\hat{\mu}_x$, or 
 to the $y$-component of the magnetic dipole operator, $\hat{m}_y$, etc.

According to the time-averaged time-dependent Hellmann-Feynman theorem\cite{Christiansen1998,langhoff1972aspects}, by defining a time-averaged quasienergy $\{Q\}_{T}$
(over the period $T$)
    \begin{equation}
        \{Q\}_{T}=\frac{1}{T}\int_{-T/2}^{T/2} \bra{0(t)}(\hat{H}-i\frac{\partial}{\partial t})\ket{0(t)} dt,
    \end{equation}
and making it stationary to changes in $\ket{0(t)}$,
we arrive at a definition of time-dependent response properties as derivatives of $\{Q\}_{T}$
\begin{equation}
\{Q\}_T = E_0 + \sum_x \langle X \rangle \epsilon_x(0) + \frac{1}{2} \sum_{x,y,k} \langle\langle X; Y\rangle\rangle_{\omega_k} \epsilon_y(\omega_k)\delta(\omega_0 + \omega_k)
\end{equation}
where
    \begin{equation}
        \langle X \rangle=\frac{d \{Q\}_{T}}{d\epsilon_{x}(0)}
    \end{equation}
corresponds to an expectation value and
    \begin{equation}
        \langle\langle X; Y \rangle\rangle_{\omega_{k}}=\frac{d^{2} \{Q\}_{T}}{d\epsilon_{x}(\omega_{0})d\epsilon_{y}(\omega_{k})}, \qquad \omega_0 = -\omega_k
    \end{equation}
to linear response properties. 


\subsection{Parametrization of the time-dependent wave-function}

In the following, we shall be concerned with wavefunctions based on an exponential parametrization of the ground state wavefunction such as the coupled cluster expansion,
\begin{equation}
    \ket{0} = e^{T_0}\ket{R} = \ketCC
\end{equation}
in which $\ket{R}$ denotes the reference state, typically the Hartree-Fock wavefunction, and $T_0$ is the time-independent cluster operator, here restricted to single ($\nu_1$) and double ($\nu_2$) excitations
\begin{equation}
T_0 =  T_1 + T_2 = \sum_{\nu_1} t_{\nu_1} \hat{\tau}_{\nu_{1}} + \sum_{\nu_2} t_{\nu_2} \hat{\tau}_{\nu_{2}} = \sum_{ai} t_i^a \{a^\dagger_a a_i\} + \frac{1}{4}  \sum_{abij} t^{ab}_{ij}  \{a^\dagger_a a^\dagger_b a_ j a_i\}
\end{equation}
with  $a,b$ indicating {particle lines} and $i,j$ hole lines, respectively~\cite{crawford2007introduction}, and $\nu_1, \nu_2$ representing excited configurations with respect to the reference ($\nu_1 \leftrightarrow \ket{\nu^a_i}= \{a^\dagger_a a_i\}\ket{R}, \nu_2 \leftrightarrow\ket{\nu^{ab}_{ij}} = \{a^\dagger_a a^\dagger_b a_ j a_i\}\ket{R}$); in the following, we shall sometimes omit explicit excitation ranks and particle/hole labels and instead employ the shorthand notation $\mu, \nu$ to denote excited determinants.

As suggested by~\citet{pawlowski_molecular_2015}, the time-dependent wave-function $\ket{0(t)}$ can be parametrized in a general manner as :
\begin{equation}
    \ket{0(t,\epsilon_{x})} = e^{B_{0}} e^{B(t,\epsilon_{x})}\ket{R},
\end{equation}
where $e^{B_{0}}$ and $e^{B(t,\epsilon_{x})}$ define the parametrization of the time-independent and time-dependent wavefunctions associated with perturbation $X$ with perturbation strength $\epsilon_{x}$, respectively. In the case of coupled cluster wavefunctions, $B_{0} = T_0$,
and the choice to be made is that of the parametrization of the time-dependent part. If the exponential parametrization is retained, we have the CC-CC model (more commonly known as LR-CC), whereas for a linearized version we have the CC-CI model (also referred to as EOM-CC)
\begin{equation}
e^{B(t,X)} \simeq 1 + B(t,\epsilon_{x}) = 1 +  \sum_{ai} t_i^a (t,\epsilon_{x}) \{a^\dagger_a a_i\} + \frac{1}{4}  \sum_{abij} t^{ab}_{ij}(t,\epsilon_{x})  \{a^\dagger_a a^\dagger_b a_ j a_i\}.
\end{equation}

\subsection{The coupled cluster linear response function}

As in the time-independent case, the non-variational nature of the coupled cluster method requires that we define a second-order quasienergy Lagrangian 

        \begin{equation}
        \{L\}_{T} = \{Q\}_{T} + \sum_{\mu}\bar{t}^{(0)}_{\mu}\{\bra{\bar{\mu}}e^{-B^{(1)}(t,\epsilon_{x})}(H-i\frac{\partial}{\partial t})e^{B^{(1)}(t,\epsilon_x)}\ketCC\}_{T}
    \end{equation}
in order to obtain the linear response functions. Here, $\bra{\bar{\mu}} \equiv \bra{\mu} e^{-T_0}$ and $\bar{t}^{(0)}_{\mu}$ are the Lagrange multipliers for the ground-state, obtained solving the linear system
\begin{equation}
\bar{\mathbf{t}}^{(0)}\mathbf{A}=-\boldsymbol{\eta},
\end{equation}
in which the matrix $\mathbf{A}$ is the Jacobian matrix. We note that $\mathbf{A}$ is strictly equivalent to the normal-ordered similarly transformed Hamiltonian $\bar{H}_N$
\begin{equation}
        A_{\mu\nu} \equiv (\bar{H}_N)_{\mu\nu}
        = \left[\exp(-\hat{T}_0)\hat{H}_0\exp(\hat{T}_0) - \bra{\mathrm{HF}}\hat{H}_0\ket{\mathrm{HF}}\right].
\label{RHS rsp equation}
\end{equation}
In the following, we shall use the two terms interchangeably, and for brevity drop the subscript $N$ in $\bar{H}_N$.

The linear response functions  are expressed as
    \begin{equation}
    \label{CC-CC response function}
    \begin{split}
        ^{\textrm{CC-CC}}\langle\langle X; Y \rangle\rangle_{\omega_{k}} = \frac{1}{2}C^{\pm\omega}P(X(\omega_0),Y(\omega_k))\left[\boldsymbol{\eta}^{X}+\frac{1}{2}\mathbf{F}\mathbf{t}^{X}(\omega_0)\right]\mathbf{t}^{Y}(\omega_k)
    \end{split}
    \end{equation}
for CC-CC~\cite{Christiansen1998,pawlowski_molecular_2015,coriani_molecular_2016} and    
    \begin{equation}
    \label{CC-CI response function}
        ^{\textrm{CC-CI}}\langle\langle X; Y \rangle\rangle_{\omega_{k}} =\frac{1}{2}C^{\pm\omega}P(X(\omega_0),Y(\omega_k))\left[^{EOM}\boldsymbol{\eta}^{X}\mathbf{t}^{Y}(\omega_k)-\sum_{\mu}\bar{t}_{\mu}^{(0)}t_{\mu}^{X}(\omega_0)\sum_{\nu}\bar{t}_{\nu}^{(0)}\xi^{Y}_{\nu}\right]
    \end{equation}
for CC-CI~\cite{pawlowski_molecular_2015,coriani_molecular_2016,faber2018resonant}. In the equations above, $P(X(\omega_0),Y(\omega_k))$ acts to permute perturbations $X$ and $Y$, and 
\begin{equation}
    C^{\pm\omega}f^{XY}(\omega_0,\omega_k) = f^{XY}(\omega_0,\omega_k) + f^{XY}(-\omega_0,-\omega_k)^*
\end{equation} symmetrizes the response functions with respect to simultaneous complex conjugation and inversion of the sign of the frequencies~\cite{Christiansen1998}. 

We have also implemented an alternative expression for the response function, that can be rewritten in an asymmetric form\cite{Christiansen1998}:
\begin{equation}
    ^{\textrm{CC-CC}}\langle\langle X; Y \rangle\rangle_{\omega_{k}} = \frac{1}{2}[\boldsymbol{\eta}^{X}\mathbf{t}^{Y}(\omega_{k})+\mathbf{\bar{t}}^{Y}(\omega_{k})\boldsymbol{\xi}^{X}]
\end{equation}
where $\mathbf{\bar{t}}^{Y}$ collect the derivatives of Lagrange multipliers with respect to one perturbation. The asymmetric form gives the same results as the symmetric form and has advantages in some cases such as NMR calculations\cite{von1998encyclopedia} as one needs to solve response equations for one operator (e.g.\ $Y$), but at the cost of having to solve response equations for both perturbed amplitudes and multipliers. In the properties investigated in this manuscript, the asymmetric form does not offer a clear advantage, and as such we focus on the symmetric form in the following.

To evaluate the linear response function, we need to obtain the frequency (in-)dependent first-order perturbed amplitudes $\mathbf{t}^{Y}$ by solving the corresponding first-order right-hand side response equations~\cite{Christiansen1998}:
    \begin{equation}
    \label{LR response equation}
        (\bar{\mathbf{H}}-\omega_{k}\mathbf{I})\mathbf{t}^{Y}=-\boldsymbol{\xi}^{Y}.
    \end{equation}
 with $\mathbf{I}$ as the identity matrix.

Because of the equivalence between $\mathbf{A}$ and $\bar{\mathbf{H}}$, Eq.~\eqref{LR response equation} is the same for the CC-CC and CC-CI models, and the poles of the response functions will occur at the same 
places in the two formulations. This is in line with the fact that excitation energies for CC-LR and EOM-CC are the eigenvalues of $\mathbf{A}$ or $\bar{\mathbf{H}}$ respectively. 

Here, we use the same definitions for matrices $\boldsymbol{\eta}^{Y}, \boldsymbol{\xi}^{Y}$ and $\mathbf{F}$ (the coupled cluster Hessian) as done by ~\citet{Christiansen1998}, which are listed in Table \ref{tab:matrix of CCLR}, and note that in the case of CC-CI, $\boldsymbol{\eta}^{Y}$ is replaced by ${^{\textrm{EOM}}}\boldsymbol{\eta}^{Y}$ as defined by \citet{faber2018resonant}. The detailed working equations used in our implementation are listed in the supplementary material.

\begin{table}[H]
\begin{threeparttable}

    \centering
    \setlength{\tabcolsep}{12.0mm}{
    \begin{tabular}{ccc}
    \hline
     $\boldsymbol{\eta}^{Y}$    &  & $\bra{\Lambda}[Y,\hat{\tau}_{\mu}]\ketCC$ \\
     $\boldsymbol{\xi}^{Y}$    &  & $\bra{\bar{\mu}}Y\ketCC$  \\
     $\mathbf{F}$    &  & $\left< \Lambda \left | \left[ \left[ H_0,\hat{\tau}_{\mu} \right],\hat{\tau}_{\nu} \right] \right|\mathrm{CC}\right>$ \\
     $\mathbf{A}$    &  & $\left< \bar{\mu}| [H_0,\hat{\tau}_{\mu}]|\mathrm{CC}\right>$ \\
    \hline
    \end{tabular}}
    \caption{Vectors and matrices for CC linear response function\tnote{a}}
    \label{tab:matrix of CCLR}
    \begin{tablenotes}
        \item[a] $\ket{CC}=e^{T_0}\ket{R}$ denote the regular CC reference wavefunction, and $\ket{R}$ is the reference state for the CC parametrization such as Hartree-Fock state. $\bra{\Lambda} = \bra{R} + \sum_{\mu}\bar{t}_{\mu}^{0}\bra{\bar{\mu}}$. $\bra{\bar{\mu}}=\bra{R}\hat{\tau}_{\mu}^{\dagger}e^{-T_0}\equiv\bra{\mu}e^{-T_0}$, where $\hat{\tau}_{\mu}^{\dagger}$ is the deexcitation operator, which is biorthogonal to excitation operator $\hat{\tau}_{\mu}$, satisfying $\left<R|\hat{\tau}_{\mu}^{\dagger}\hat{\tau}_{\nu}|R\right> = \delta_{\mu\nu}$.
    \end{tablenotes}
\end{threeparttable}
\end{table}

Finally, due to the fact that ExaCorr was originally designed for treating systems without symmetry and that in such a case the relativistic wave functions are complex-valued, complex algebra is used throughout. This makes the implementation of damped coupled cluster response theory relatively straightforward; it suffices, in the computation of the response function of interest (for instance the yy component of the electric dipole polarizability, ${\alpha_{yy}}(\omega_0;\omega_k)$), to set the imaginary component of the perturbing frequency $\omega_k$ to a particular inverse lifetime $\gamma$
\begin{equation}
\omega_k \equiv \omega_k + i0 \rightarrow \omega_k + i\gamma
\end{equation}
when solving the response equation\cite{norman2001near,norman2005nonlinear,coriani2012coupled}:
    \begin{equation}
    \label{LR response equation_cpp}
        (\bar{\mathbf{H}}-(\omega_{k}+i\gamma)\mathbf{I})\mathbf{t}^{Y}(\omega_{k}+i\gamma)=-\boldsymbol{\xi}^{Y},
    \end{equation}
subject to the condition that $(\omega_k + i\gamma)+(-\omega_0 -i\gamma) = 0$. We note that, while we can in principle use a different value of $\gamma$ for each $\omega_k$, in practice we will follow common usage and keep this value constant for a range of frequencies for which we shall calculate a particular response function. With that, the absorption cross-section for dipole transitions can be determined by the imaginary part of the complex electric dipole  polarizability\cite{boyd2020nonlinear} :
    \begin{equation}
        \sigma(\omega) = \frac{4\pi \omega}{c} \mathrm{Im}[\bar{\alpha}(\omega)]
        \label{eq:abs_cpp}
    \end{equation}

\section{Implementation}
\label{Implementation}

The above-mentioned algorithm has been implemented in the development version of the relativistic quantum chemistry package DIRAC\cite{saue2020dirac}  as a part of the ExaCorr code~\cite{pototschnig_implementation_2021}. Currently, the implementation allows for calculations to be carried out only using a single-node configuration. The implementation of multi-node is currently in progress and will be reported in forthcoming works. We can summarize the main computational tasks in the following four steps :
\begin{enumerate}
    \item Solve closed-shell ground state CCSD equations to obtain the $\bf{t}_1, \bf{t}_2$ amplitudes. 
    \item With $\bf{t}_1, \bf{t}_2$, construct the one and two-body intermediates, that are necessary for building the $\bar{\mathbf{H}}$ and linear response functions.
    \item Solve the linear response equation in the full single-double excitation space to obtain the first-order perturbed amplitudes for each operator-frequency combination. To avoid the explicit construction of large matrix $\bar{\mathbf{H}}$, an iterative solver is employed. 
    \item Construct the response function by combining the first-order perturbed amplitudes and the property integrals in the molecular orbital (MO) basis. 
\end{enumerate}

The first step is carried out within a Kramers-unrestricted formalism\cite{visscher1996formulation} and has been extensively discussed in prior work\cite{pototschnig_implementation_2021}. 

The intermediates in the second step consist of two sets: The first set is property-independent and is utilized to construct the $\boldsymbol{\sigma}$-vectors, which are the projection of $\bar{\mathbf{H}}$ in the trial vector space. These intermediates were previously discussed in the literature\cite{asthana_exact_2019,peng_energy-specific_2015} and an implementation of relativistic EOM-CC 
 is available in the RelCCSD module~\cite{shee2018equation} as well. We have included a rewritten version of the  $\boldsymbol{\sigma}$-vectors for EOM-CC for excitation energies (EOM-EE) in the supplementary material  (for completeness, expressions for the left EOM-EE $\boldsymbol{\sigma}$-vectors  are also given), due to our use of full tensors in this implementation. We have also corrected misprints identified in the expressions given by~\citet{shee2018equation} (the previously implemented expressions were verified and found to be correct).

In deriving the working equations, we note that for the matrix $\mathbf{F}$ it is not possible to obtain its matrix elements diagrammatically \cite{shavitt2009many}, due to the number of unconnected hole/particle-lines. However, $\mathbf{F}$ is never used by itself but rather as the vector-matrix product  $\mathbf{t}^{X}\mathbf{F}$, in analogy to the $\boldsymbol{\sigma}$-vector expressions for the eigenvalue and response equations. Apart from being readily expressed diagrammaticaly, dealing with the vector-matrix product is computationally advantageous as it reduces storage requirements.   

Moreover, as~\citet{faber2018resonant} suggested, we can avoid computing the \textbf{F} matrix in CC-CI implementation by computing the $^{\textrm{EOM}}\boldsymbol{\eta}^{X}$, which is easily accomplished by modifying the existing $\boldsymbol{\eta}^{X}$ routine (see working equation in the supplementary material).

In the current implementation, the set of all property-related intermediates is computed on the MO basis, but as the property integrals are first generated in the atomic orbital (AO) basis in DIRAC, it is necessary to transform all desired property integrals from AO to MO prior to the response calculations. All the intermediates and the property integrals in MO basis are therefore stored as tensor objects according to the definition of the TAL-SH library\cite{githubRepo2023}, so that they can be efficiently employed in constructing the elements of $\boldsymbol{\xi}^{X}, \boldsymbol{\eta}^{X}$ and $^{\textrm{EOM}}\boldsymbol{\eta}^{X}$.

In the supplemental information, we will focus on discussing the third step, which involves solving the first-order response equation. There are different algorithms to solve linear equations such as direct inversion of the iterative subspace (DIIS)\cite{scuseria_accelerating_1986,hattig_cc2_2000,nanda_static_2016}, as well as the Lanczos-chain\cite{hansen_lanczos-chain_2010,coriani_asymmetric-lanczos-chain-driven_2012} and Davidson~\cite{olsen1988solution} schemes. In the current work, we utilized the latter, which required minor modifications with respect to evaluating the eigenvalues and eigenvectors of $\bar{\mathbf{H}}$ in our EOM-CC implementation. We also note that the default algorithm in ExaCorr to solve for the unperturbed amplitudes $\boldsymbol{t}$ was recently changed to Conjugate Residual with OPtimal trial vectors (CROP)\cite{Ziolkowski2008,Ettenhuber2015} as this reduces the memory requirements in this stage of the calculation. 

One particular difference between our implementation and the one by~\citet{shee2018equation} is that the ExaTENSOR and TAL-SH libraries, for reasons of scalability and generality, do not enforce triangularity or the (anti)symmetry of tensors with respect to exchange of pairs of indices. Consequently, and in contrast to the prior implementation, beyond rank-2 tensors antisymmetry needs to be enforced in order to ensure that at all times we satisfy the underlying fermionic nature of the problem.

For example, in the generation of trial vectors for the ${\bf{t}}^x_2$ amplitudes (or ${\bf{r}}_2$ in EOM-CC), this means that we pick out a unique element $\ket{^{ab}_{ij}}$, where $a>b$ and $i>j$, generate the permutations and antisymmetrize them. During Davidson iterations we also ensure that the trial vectors remain antisymmetric during the Gram-Schmidt orthonormalization process, as we found that if explicit antisymmetrization is not carried out, numerical noise may lead to loss of the antisymmetry in new vectors during iterations.

With respect to the choice of starting vectors, differently from the eigenvalue case in which the pivoting was done on the basis of the value of the diagonal of $\bar{\mathbf{H}}$ (see~\citet{shee2018equation} for details), for linear systems, the pivoting is done on the basis of the magnitude of the property gradient $\boldsymbol{\xi}^{X}$ vector elements (from highest to lowest), in order to avoid selecting initial vectors with zero norm.

\section{Computational details}
\label{ComputDetails}
All coupled cluster linear-response calculations were carried out with a development version of the DIRAC code\cite{saue2020dirac,DIRAC23}, employing the uncontracted singly-augmented valence double zeta Dyall basis set s-aug-dyall.v2z for the heavy elements (Zn~\cite{dyall2009relativistic}, Cd~\cite{dyall2009relativistic}, Hg~\cite{dyall2009relativistic}, Cs~\cite{dyall2009relativistic}, I~\cite{dyall2022diffuse,dyall2006relativistic},  T~e\cite{dyall2022diffuse,dyall2006relativistic}), and a similar uncontracted Dunning basis set aug-cc-pVDZ for the light elements (H~\cite{kendall1992electron}, Li~\cite{prascher2011gaussian}, Na~\cite{prascher2011gaussian}, K~\cite{hill2017gaussian}, F~\cite{kendall1992electron}, Cl~\cite{woon1993gaussian},  O~\cite{kendall1992electron}, S~\cite{woon1993gaussian}, Se~\cite{wilson1999gaussian}, Br~\cite{wilson1999gaussian}). In most calculations, we utilized the exact two-component (X2C)\cite{iliavs2007infinite} relativistic Hamiltonian, where the spin-orbit operator takes the form of an effective one-electron operator. The screening of the nuclear charges is in the version that we used approximated by an atomic mean field\cite{Schimmelpfennig.Gropen.1998f8}. To show the effect of relativity explicitly we also provide results using the non-relativistic Hamiltonian\cite{levy1967nonrelativistic,visscher2000approximate}. We furthermore show some results with  the spin-free X2C Hamiltonian in which spin-orbit coupling terms are identified by transforming to the modified Dirac representation\cite{dyall1994exact} and eliminated prior to defining the X2C transformation and Hamiltonian. To study the effect of electron correlation, we performed linear-response calculations based on mean-field methods like Hartree-Fock (HF)\cite{saue2003linear} as well as density-functional theory\cite{salek2005linear} (especially with the B3LYP\cite{becke1993new} density functional approximation). The relativistic and non-relativistic calculations have been carried out with the Gaussian type\cite{visscher1997dirac} and  point charge nucleus model, respectively. 

In our calculations, we have profited from the components of an ongoing implementation in ExaCorr of the Cholesky-decomposition\cite{beebe1977simplifications,koch2003reduced,aquilante2007low} approach to reduce the memory footprint of our calculations in the step to transform two-electron integrals from AO to MO basis, by avoiding the storage in memory of the whole AO basis two-electron integral tensor. The Cholesky vectors (generated with a  conservative threshold of 10$^{-9}$, as to retain most of them) are then used to explicitly form all six two-electron integral classes needed by the coupled cluster method. In a subsequent publication~\cite{Pototschnig_CD_2023}  we shall address the use of Cholesky vectors directly in the coupled cluster implementation of ground and excited state properties. 

The molecular structures employed in all calculations are taken from the literature: In case of the diatomic molecules, from~\citet{huber1979constants} for HX(X=F, Cl, Br, I), I$_{2}$, ICl, from \citet{hessel1971experimental} for NaLi, and from \citet{ferber2008ground} for CsK. The internuclear distances employed are thus H-F(0.91680 \AA), H-Cl(1.27455 \AA), H-Br(1.41443 \AA), H-I(1.60916 \AA), Cl-I(2.32087 \AA), I-I (2.6663 \AA), Li-Na(2.81 \AA), and K-Cs(4.285 \AA). For the chiral molecules H$_{2}$Y$_{2}$(Y=O, S, Se, Te), the Y-Y bond length, H-Y bond length, and H-Y-Y bond angle are taken from Table I of~\citet{laerdahl1999fully} and the dihedral angle is kept fixed at 45 degrees.

The size of the correlated virtual spinor spaces in the coupled cluster 
calculations is truncated by discarding spinors with energies above 5 a.u. For the IIB atoms, we correlate both semi-core and valence electrons for Zn(3d,4s), Cd(4d,5s), Hg(5d,6s), respectively. In the polarizability and optical rotation calculations of molecular systems, we correlate only valence electrons. In the spin-spin coupling calculations, which are known to be more sensitive to core relaxation and correlation, we correlate all occupied and virtual orbitals. 

All optical rotation calculations (HF, DFT, CC) employed a common gauge origin, set to the origin of the coordinate system, chosen at the midpoint of the bond between the two chalcogen atoms, which nearly coincides with the systems' center of mass. The atomic coordinates for each system under consideration as well as further details on the calculations (position of center of mass etc.) are provided respectively as XYZ and output files in the dataset associated with this manuscript (see ``Supporting Information Available'').

\section{Sample applications}
\label{Results}

\subsection{Polarizability of IIB atoms}
We begin the discussion by analyzing the obtained results for the polarizability of the Zn, Cd, and Hg atoms and present in Table \ref{tab:staticIIB} the static polarizability of these atoms, calculated by different methods. A comparison of the first three rows of this Table shows the growing influence of relativity on the static polarizability from Zn to Hg. For example, the relativistic HF value for Hg (44.82 a.u.) is nearly half of its nonrelativistic counterpart (81.05 a.u.). This is mainly due to the strong relativistic contraction of the $6s$-shell.

In contrast, the effect of electron correlation at CCSD level is rather constant for these elements ( $-$10.15 a.u. for Zn, $-$15.39 a.u, and $-$9.57 a.u, for Hg). 

For Zn, electron correlation primarily accounts for the discrepancy between HF and the experimental results. However, for Cd and especially Hg, the inclusion of relativity is crucial. The above-mentioned contraction of the valence s-shell reduces the magnitude of the polarizability, whereas spin-orbit coupling (SOC) becomes increasingly important by enabling spin-forbidden transitions. We will discuss this consequence of relativity in greater depth when looking at the frequency-dependence of the polarizability in the next section. 

An error of approximately 1-2 a.u. 
remains between our relativistic CCSD results and the experimental values. To locate the source of this error, we performed CCSD(T) calculations using the finite-field method, since the analytic gradient is not yet available for CCSD(T) in DIRAC. In these CCSD(T) finite-field calculations, we used an external electric strength of 0.005 a.u, which is sufficiently large to avoid numerical issues and small enough to remain in the linear regime. 

From a comparison between the results of CCSD and CCSD(T), it is evident that the inclusion of the triple excitations indeed enhances the accuracy: from 95.99\% to 99.84\% for Zn, from 98.44\% to 99.71\% for Cd, and from 96.20\% to 97.95\% for Hg, respectively. In the results of CCSD(T), we also performed a calculation in which all virtual orbitals were used (so without energy truncation), but this did not significantly affect the results. Upon using the valence triple-zeta basis set s-aug-dyall.v3z, the Hg results improve and come close to the experimental error bar.   

    \begin{table}[H]
        \begin{threeparttable}
            \center
            \caption{\label{tab:staticIIB}Static polarizability (a.u.) of IIB atoms calculated with the X2C Hamiltonian}
            \setlength{\tabcolsep}{5.5mm}
            {
            \begin{tabular}{cccc}
            & Zn    & Cd   & Hg \\
            \hline
            NR-HF\tnote{a}  & 53.88 & 76.01& 81.05\\
            SF-HF\tnote{b}  & 50.58 & 63.65& 44.90\\
            HF  & 50.57 & 63.64& 44.82\\
            SF-CCSD\tnote{b} &40.42 &48.28 &35.35\\
            CCSD & 40.42 & 48.25& 35.25\\
            CCSD(T) & 38.86 & 47.64& 34.62\\
            CCSD(T) (all virtual dz)& 38.80 & 47.69& 34.66\\
            CCSD(T) (tz)& 38.86 & 46.64& 34.27\\
            Exp& 38.8$\pm$0.80\cite{goebel1996theoretical} & 47.5$\pm$2\cite{hohm2022dipole}& 33.91$\pm$0.34\cite{goebel1996dipole} \\
            \hline    
            \end{tabular}
            }
            \begin{tablenotes}
                \item [a] Nonrelativistic calculation with the Levy-Leblond Hamiltonian
                \item [b] Scalar relativistic calculation with the spin-free\cite{dyall1994exact} X2C Hamiltonian
            \end{tablenotes}
        \end{threeparttable}
    \end{table}

\begin{figure}[H]
    \centering
    \fbox{\includegraphics[width = .5\linewidth,height=6.0cm]{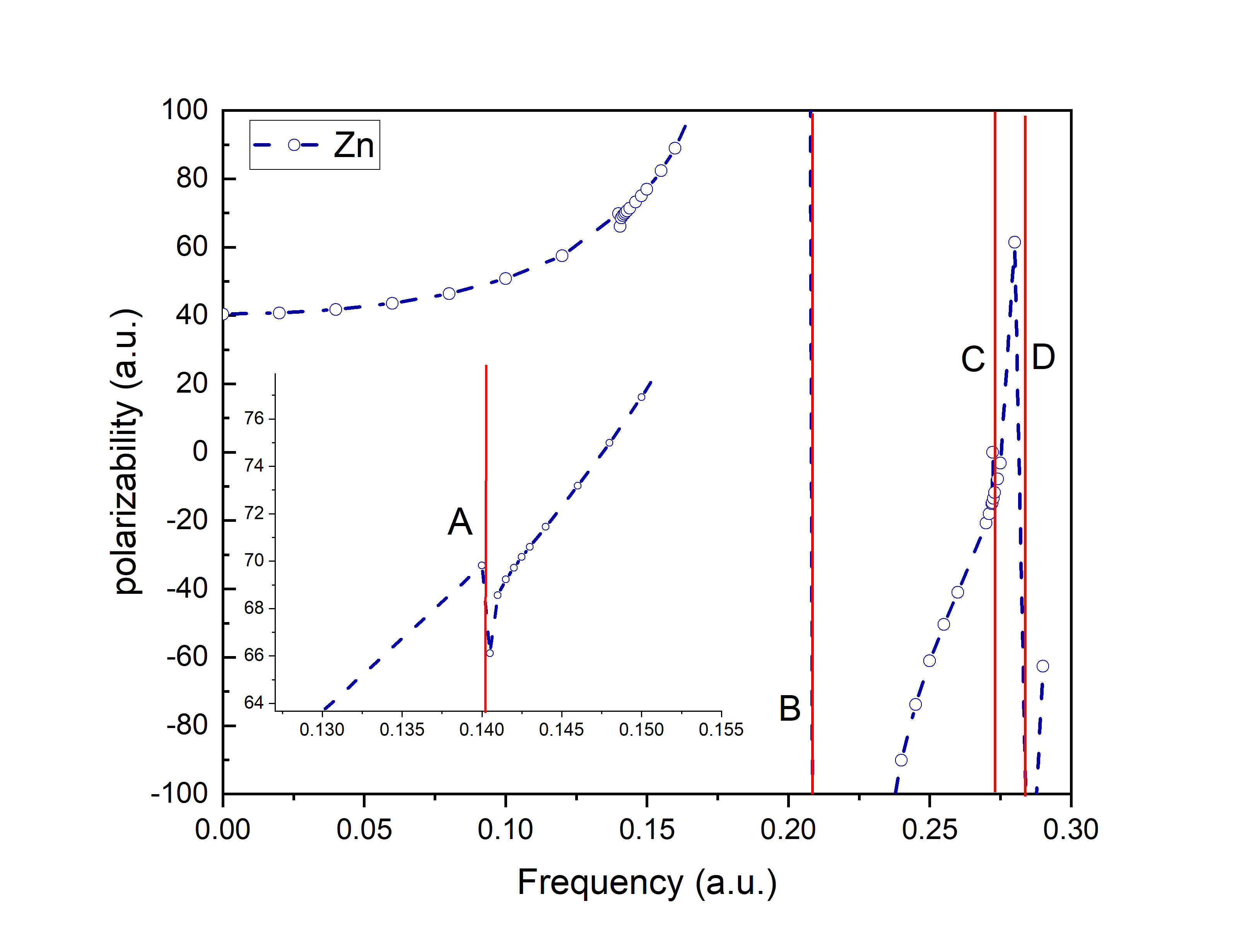}
          \includegraphics[width = .49\linewidth,height=6.0cm]{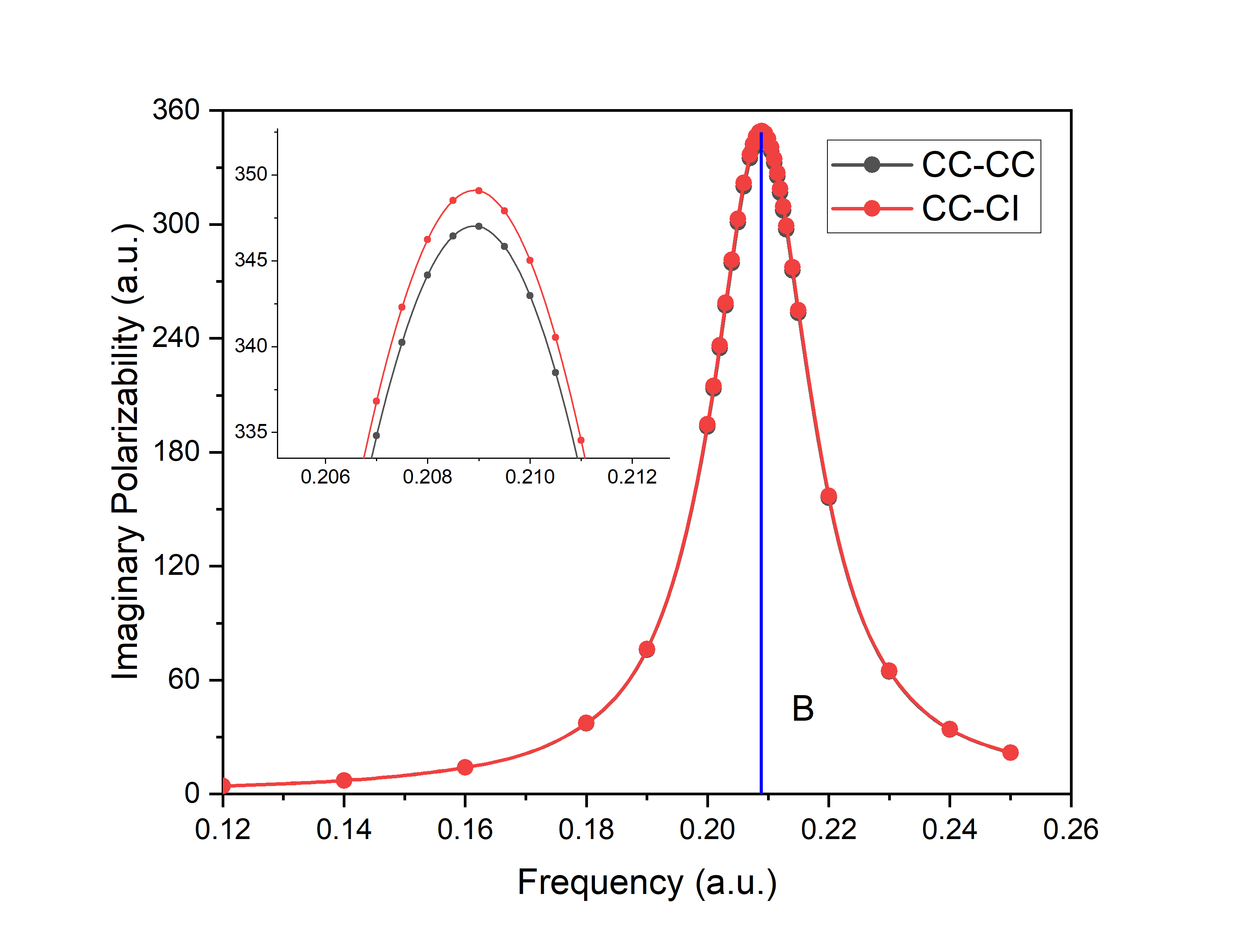}}
    \fbox{\includegraphics[width = .5\linewidth,height=6.0cm]{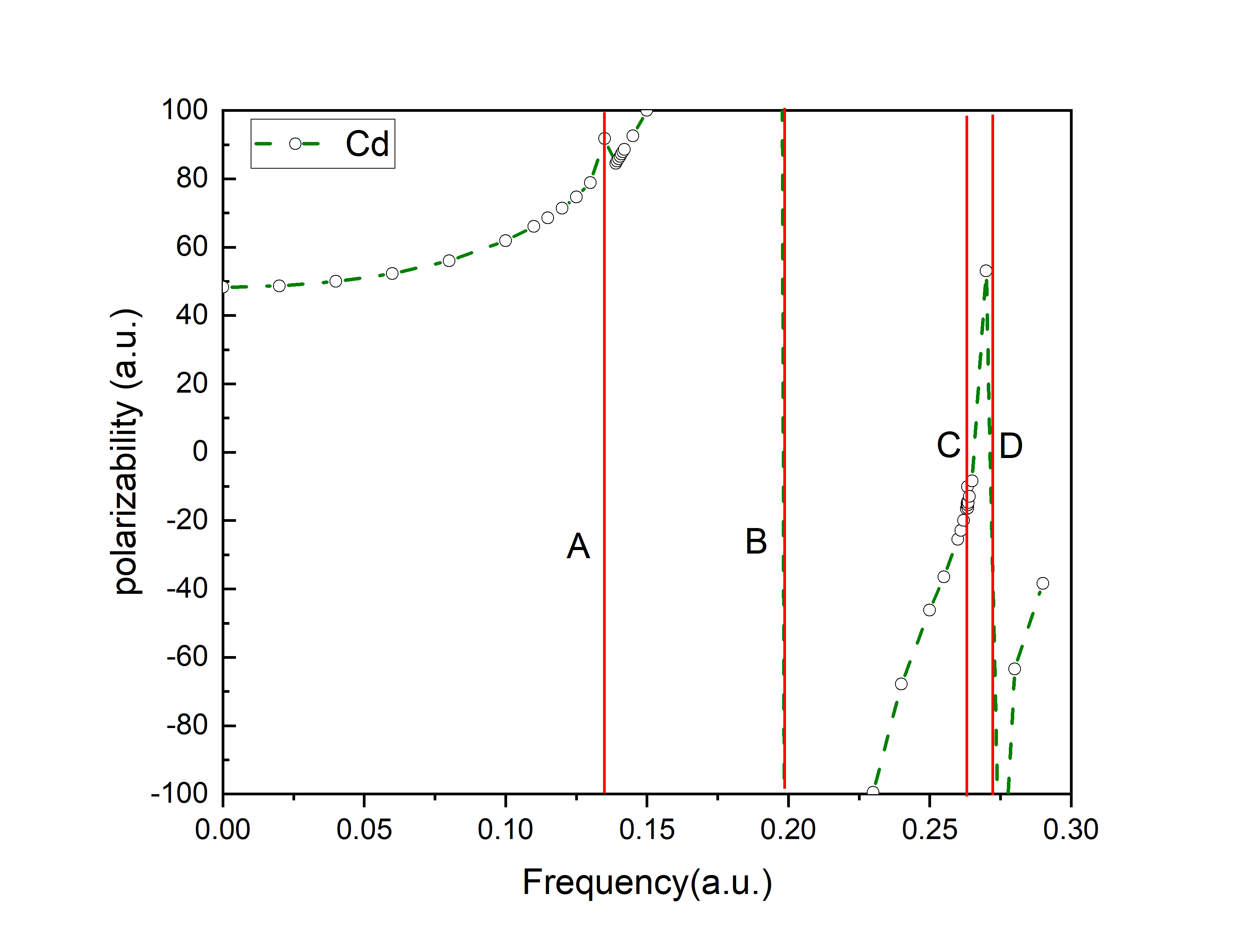}
    \includegraphics[width = .49\linewidth,height=6.0cm]{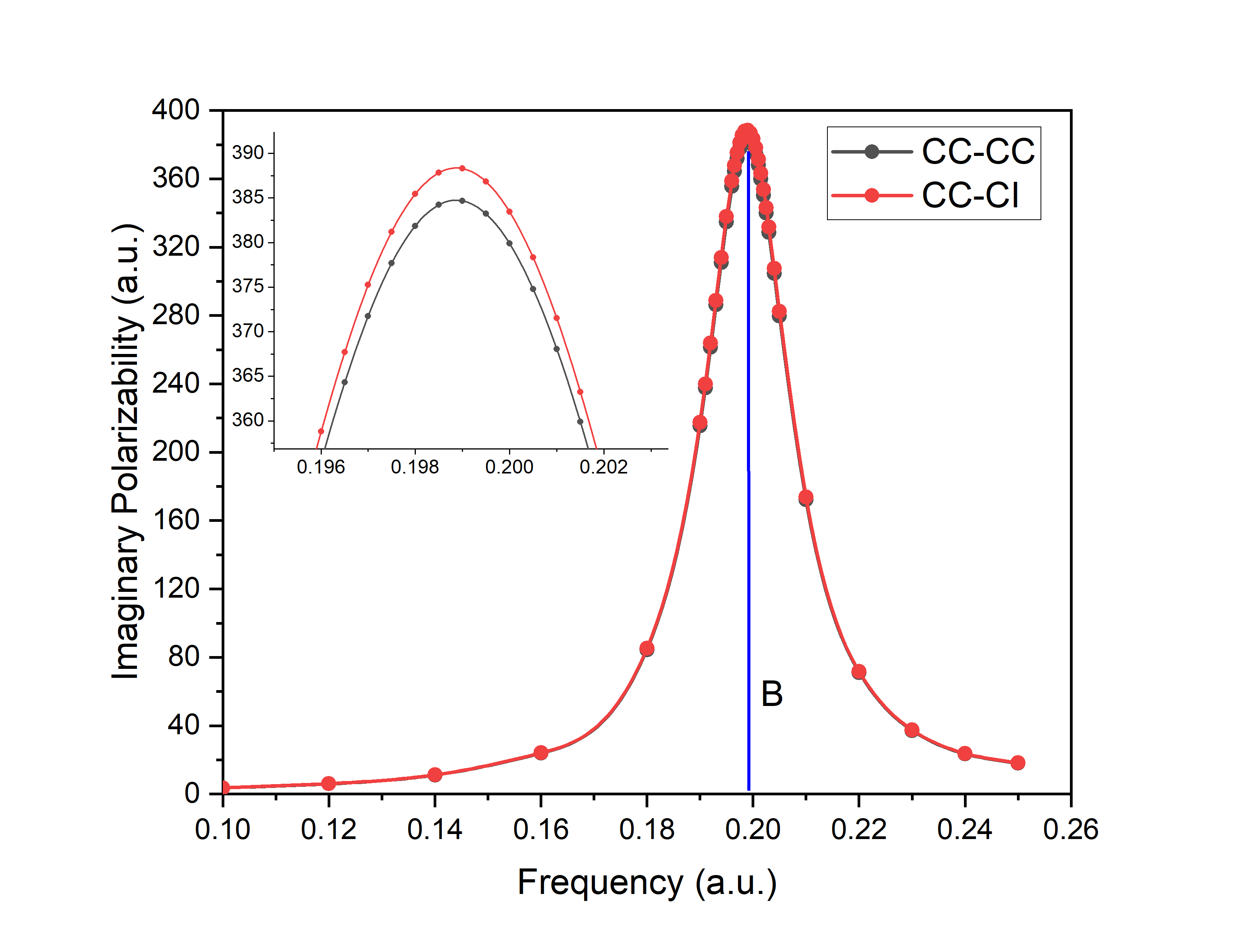}}
    \fbox{\includegraphics[width = .5\linewidth,height=6.0cm]{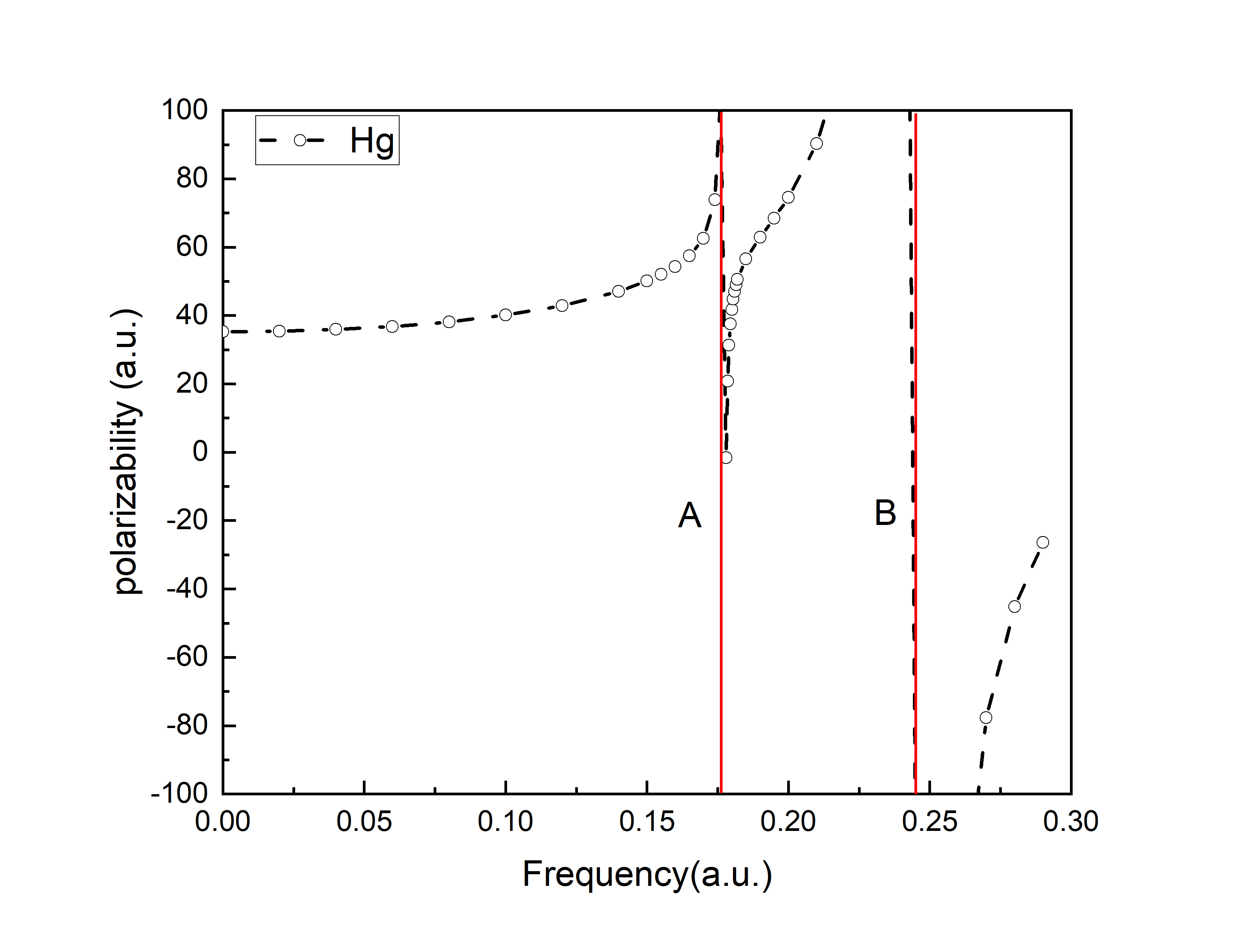}
    \includegraphics[width = .49\linewidth,height=6.0cm]{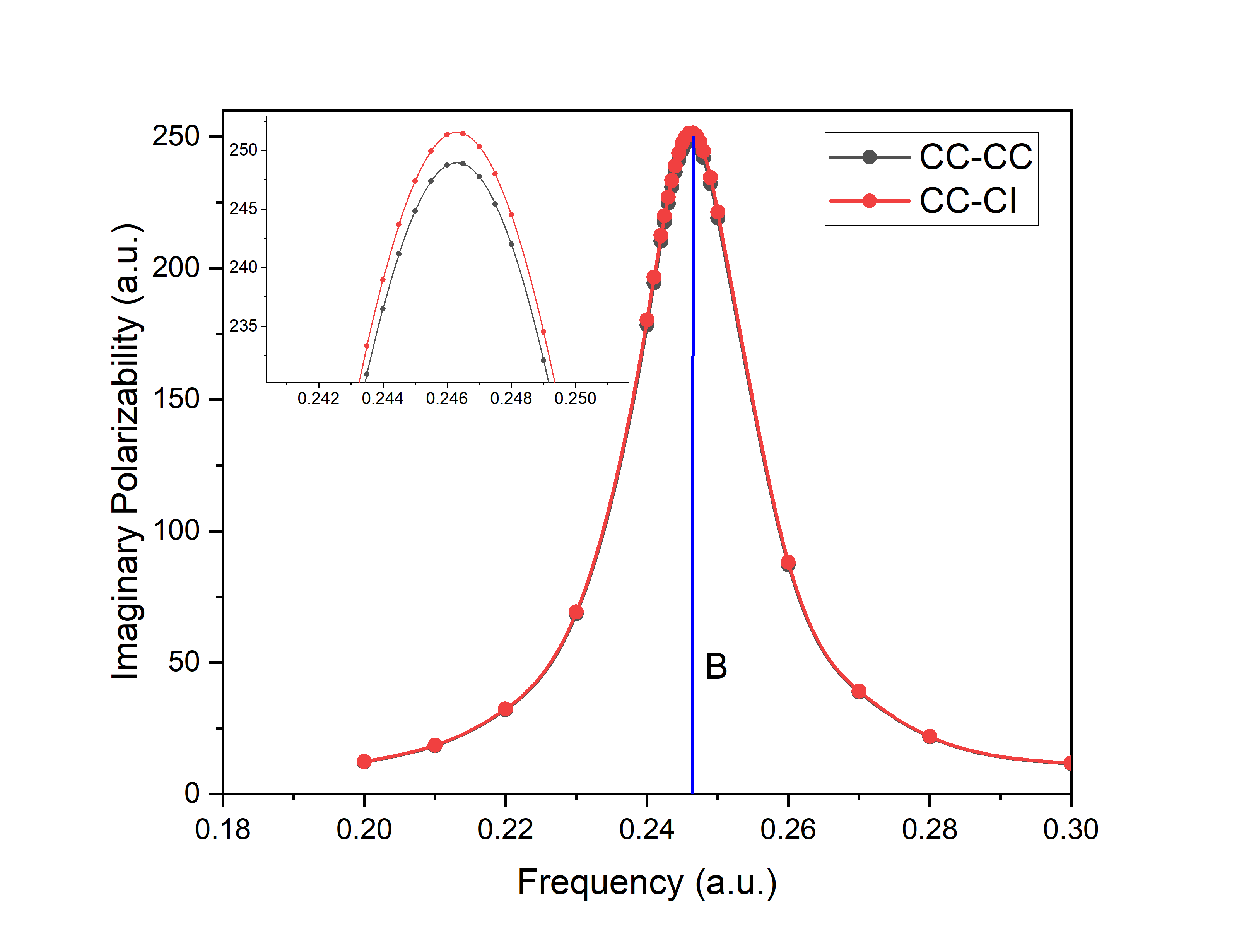}}
    \caption{Frequency dependent polarizability of Zn(top), Cd(middle), and Hg (bottom). The left figures show the real (undamped) polarizability with frequency ranging from 0.0 to 0.30 a.u. The right figures, with insets to zoom in on the peak maxima, show the simulated spectra focusing on the \textbf{B} transition, and are obtained from the imaginary part of the complex (damped) polarizability with an imaginary component of the frequency $\gamma = 0.01$ a.u. The red vertical lines are the EOM excitation energy: For \textbf{Zn}: A($^{3}P_{1}^{o}(4s^{1}4p^{1})$0.1403 a.u.); B($^{1}P_{1}^{o}(4s^{1}4p^{1})$0.2089 a.u.); C($^{3}P_{1}^{o}(4s^{1}5p^{1})$0.2723 a.u.); D($^{1}P_{1}^{o}(4s^{1}5p^{1})$0.2828 a.u.). For \textbf{Cd}:  A($^{3}P_{1}^{o}(5s^{1}5p^{1})$0.1356 a.u.); B($^{1}P_{1}^{o}(5s^{1}5p^{1})$0.1989 a.u.); C($^{3}P_{1}^{o}(5s^{1}6p^{1})$0.2634 a.u.); D($^{1}P_{1}^{o}(5s^{1}6p^{1})$0.2728 a.u.). For \textbf{Hg}: A($^{3}P_{1}^{o}(6s^{1}6p^{1})$0.1771 a.u.); B($^{1}P_{1}^{o}(6s^{1}6p^{1})$0.2463 a.u.).}
    \label{fig:fre-dep-iib}
\end{figure}

We now turn to the frequency-dependence of the polarizability and look at the effect of SOC. In Figure \ref{fig:fre-dep-iib}, the frequency-range from 0.0 to 0.30 a.u. is displayed. The  singularities at the frequencies of spin-allowed transition $^1S_0 \rightarrow ^1P_1 (ns\rightarrow np)$ dominate these curves, while the non-relativistically spin-forbidden transition to the $^3P$ state is clearly visible for Hg and, after zooming in on the transition energy, also already for Zn. Calculating and plotting the polarizability~\cite{kauczor2013communication}
\begin{equation}
    \alpha_{\alpha\beta}(\omega) = - \sum_{n} \left[ \frac{\bra{0}\hat{\mu}_{\alpha}\ket{n}\bra{n}\hat{\mu}_{\beta}\ket{0}}{E_{n} - \omega} + \frac{\bra{0}\hat{\mu}_{\beta}\ket{n}\bra{n}\hat{\mu}_{\alpha}\ket{0}}{E_{n}+\omega} \right]
    \label{fre-den-pol}
\end{equation}
over a range of frequencies $\{\omega\}$ implicitly shows all excitation energies $E_n$  in the associated energy range. However, when interested in the values of these energies it is of course more efficient to directly solve for the poles by diagonalizing $\bar{\mathbf{H}}$. To check the correctness of the implemented solvers, we therefore compared the linear response and EOM-EE results employing the same Hamiltonian and basis set. The resulting excitation energies are depicted in Fig \ref{fig:fre-dep-iib} with red lines and do indeed precisely align with the pole locations in the polarizability curves. 

Looking at the low-lying parity-allowed ($ns \rightarrow mp$) transitions in the studied frequency-range, for Zn and Cd we find two $ns \rightarrow (n)p$ transitions (A and B, respectively spin-forbidden ${}^{1}S_{0} \rightarrow {}^{3}P^{o}_{1}$ and spin-allowed ${}^{1}S_{0}\rightarrow{}^{1}P^{o}_{1}$ transitions) and two $ns \rightarrow (n+1)p$ transitions (C and D, similarly spin-forbidden and spin-allowed transitions). For Hg on the other hand, only A and B are within the studied frequency-range, with C and D coming at higher energies and therefore not observed.

On the right side of Fig~\ref{fig:fre-dep-iib}, we also show a simulated spectrum of the first spin-allowed transition $^{1}S_{0} \to {^{1}}P^{o}_{1}$ by calculating the damped linear response function for both CC-CC and CC-CI models. While CC-CI is an approximation of the CC-CC model, we note that the CC-CI curve exhibits a shape very similar to that of CC-CC curves, in that both are Lorentzian-type line shapes and share the exact same peak location since they solve the same response equation as demonstrated in the Eq.~\eqref{LR response equation}. The CC-CI model spectrum shows only a minor difference in the peak height with a relative error of about 1\%. To verify the implementation of the complex polarizability, we pay particular attention to the peak value of the spectrum of B atomic transition. Around the pole of the transitions we are investigating, the stationary point in the curves should be well approximated by the norm of the transition dipole moment divided by $\gamma$,
\begin{align}
    \mathrm{Im} \left(\alpha_{\alpha\beta}(\omega)\right) 
 &= \mathrm{Im} \left(- \sum_{n} \left[ \frac{\bra{0}\hat{\mu}_{\alpha}\ket{n}\bra{n}\hat{\mu}_{\beta}\ket{0}}{E_{n} - \omega -i\gamma} + \frac{\bra{0}\hat{\mu}_{\beta}\ket{n}\bra{n}\hat{\mu}_{\alpha}\ket{0}}{E_{n}+\omega+i\gamma} \right]\right)
 \nonumber \\
    \label{complex-fre-den-pol} 
    &\approx \mathrm{Im}\left(\frac{\bra{0}\hat{\mu}_{\alpha}\ket{n}\bra{n}\hat{\mu}_{\beta}\ket{0}}{i\gamma}\right) 
\end{align}
In the current work,  we set the imaginary component of the frequency $\gamma$ as 0.01 a.u. for all three atoms. Even though the EOMCC transition dipole moment is not yet available in DIRAC, 
we can still compare the intensity ratios (Zn:Cd:Hg) between our results (1.39:1.55:1.0) and the values derived from the experimental lifetimes\cite{lurio1964lifetime_Zn,lurio1964lifetime,pinnington1988lifetime} (1.48:1.54:1.0). It is noteworthy that our results qualitatively mirror the experimental trend. The small difference in ratios likely stems from the omission of higher-order correlation and the quality of the basis set used. In supplementary information, we simulate the spectrum of BH molecules with damped CC-CC and find our results exactly reproduce the DALTON results. 

\subsection{Polarizability of Molecules}
As our implementation is mainly intended for molecular systems, we will now look at results for molecular polarizabilities which may have up to three distinct values upon diagonalizing the polarizability tensor. For diatomic and other symmetric molecules it is sufficient to consider the mean dipole polarizability $\alpha(\omega)$ and the anisotropy $\Delta\alpha(\omega)$:
    \begin{align}
        \alpha(\omega) = & \frac{1}{3}\left(\alpha_{zz}(\omega)+\alpha_{xx}(\omega)+\alpha_{yy}(\omega)\right)\\
        \Delta\alpha(\omega) = & \alpha_{zz}(\omega)-\frac{1}{2}\left(\alpha_{xx}(\omega)+\alpha_{yy}(\omega)\right),
    \end{align}

\noindent where $z$ is the molecular symmetry axis.
In Table \ref{tab:static polarizability of diatomics}, we list the static mean and anisotropic polarizability of hydrogen halides and alkali-metal diatomic molecules assessed by HF, B3LYP, and CC models with both relativistic (X2C) and nonrelativistic Hamiltonians, and the corresponding experimental values as well. Unless otherwise specified, 'CC' refers to 'CC-CC-LR'.

The HF results deviate from the experimental value for both the mean and anisotropic polarizability and the impact of the relativistic effect increases as we move from lighter to heavier molecules. For example, the relativistic correction is nearly zero for hydrogen fluoride but amounts to 1.3 a.u. for I$_{2}$. In the case of the CsK molecule, the relativistic correction at the HF level is 31 a.u, emphasizing the necessity of  considering the relativistic effect in the calculation of heavy elements. For this molecule, the effect of relativity may again be rationalized in terms of contraction of the outermost valence $s$-orbitals, in particular that of Cs, which reduces the polarizability, similar to what we observed in the Hg atom. 

Apart from relativity, another source of discrepancy between HF and  experiment lies in the importance of electron correlation. Electron correlation is modeled in DFT by the B3LYP functional, and explicitly calculated in the CC models. From the results, it is evident that in both cases the electron correlation and the relativistic correction are not strictly additive. For example, computed with CC the relativistic corrections for I$_{2}$ and CsK are 0.91 a.u. and -26 a.u, respectively, while they are 1.30 a.u. and -31 a.u. when computed with HF.

The B3LYP calculations yield much better values than HF for both the relativistic and nonrelativistic Hamiltonian. The relativistic correction on the B3LYP model is similar in magnitude as found for HF, but with a different sign. For instance, the relativistic correction of anisotropic polarizability for CsK for HF is +9 a.u while it is $-11$ a.u. in B3LYP. 
For the halides, 
the B3LYP calculations yield values that are close to the CC results and arewithin or only slightly outside the experimental error bars for both isotropic and anisotropic polarizability. However, for the alkali-metal diatomic molecules NaLi and CsK, the B3LYP values significantly deviate from the experimental value.

The CC results are close to the experimental data for both halide and alkali-metal molecules. We have also tried using the triple-zeta basis set for CC on three light hydrogen halide molecules (HF, HCl HBr) to reduce the error and indeed observe an improvement of CC values which then fall within the experiment error bar for the isotropic polarizability. Getting the smaller anisotropic polarizability agree with experimental data is more demanding on the model and may require addition of more diffuse functions and/or the inclusion of the triple excitations. 

One may note for the anisotropic polarizability of HI the considerable deviation of all three theoretical values (around 2 a.u.) from the experimental value of 11.4 a.u. that was determined in 1940 by Denbigh\cite{denbigh1940polarisabilities}. Curiously, this value appears to have not been re-evaluated since then, while the isotropic polarizabilities of HCl and HBr, that were also reported by Denbigh, were later estimated to be significantly lower by Kumar and Meath\cite{kumar1985integrated}. The anisotropy of HBr that was given as 6.14 a.u. by Denbigh was adjusted to just 1.7 a.u. by Pinkham and Jones\cite{pinkham2008extracting}, but we could not find a similar re-evaluation of the anistropic polarizability of HI on basis of experimental data in the literature. This discrepancy between theory and the old experimental value for the anisotropy was also noted  in theoretical work of~\citet{maroulis2000dipole} and~\citet{iliavs2003electric}. \citeauthor{iliavs2003electric} used relativistic CCSD(T) and included vibrational corrections on both dipole moment and static polarizability and found their results to be significantly lower than the experimental value: their anisotropic polarizability was 2.33$\pm$0.05 a.u, which agrees well with the current relativistic CC linear response value of 2.51 a.u. While their suggestion that also the experimental value of the dipole moment could be inaccurate could not be not sustained\cite{van2004theoretical, Li_HIReference_2013} we agree that the discrepancy between theory and experiment for the anisotropic polarizability is likely due to an inaccuracy in the experimental value. Nonetheless, it would be nice to put more firm error bars on the theoretical value as well by employing a larger basis set, including g and h functions. This was not feasible with our current implementation due to  memory constraints related to the use of a single compute node.

        \begin{table}[H]
        \begin{threeparttable}        
            \center
            \caption{Static dipole polarizability (a.u.) of diatomic molecules}\label{tab:static polarizability of diatomics}

        \setlength{\tabcolsep}{2.5mm}{
            \begin{tabular}{cccccccccc}
                &HF\tnote{a}  &HF\tnote{b}& B3LYP\tnote{a} & B3LYP\tnote{b} & B3LYP\tnote{c} &  CC\tnote{a} & CC\tnote{b}  & CC\tnote{c} & Exp \\
            \hline
            \multicolumn{10}{c}{Mean dipole polarizability}\\
            \hline
            HF  &4.40 &4.40&5.11 &5.12 &5.57  &5.04 &5.05 &5.52 &5.60$\pm$0.10\cite{kumar1985integrated}\\
            HCl &15.51 &15.54&16.34 &16.38 &17.53 &16.06 &16.09 &17.14 &17.39$\pm$0.20\cite{kumar1985integrated}\\
            HBr &21.86 &21.90&22.85 &22.94 &24.43 &22.52 &22.58 &24.02 &23.74$\pm$0.50\cite{kumar1985integrated}\\
            HI  &33.62 &33.50&34.63 &34.69 &36.71&34.36 &34.30 & &35.30$\pm$0.50\cite{cuthbertson1914info}\\
            ICl &46.48 &46.52&47.53 &47.66 &49.80 &47.48 &47.59 && 43.8$\pm$4.4\cite{swift1988dispersion}\\
            I$_{2}$ &67.90 &69.20&68.72 &69.92 &72.55 &68.81 &69.72 & &69.7$\pm$1.8\cite{maroulis1997electrooptical}\\
            NaLi &231 &230 &210 &209 &210 &240 &240 &239 &263$\pm$20\cite{antoine1999static}\\
            CsK &723 &692 &581 &548 &549 &637\tnote{d} & 611\tnote{d}& &601$\pm$44\cite{tarnovsky1993measurements}\\
            \hline
            \multicolumn{10}{c}{Anisotropic dipole polarizability}\\
            \hline
            HF  &1.79 &1.79 &1.91 &1.91 &1.46 &1.96 &1.96 &1.45 &1.62\cite{muenter1972polarizability}\\
            HCl &2.35 &2.34 &2.18 &2.16 &1.66 &2.39 &2.38 &1.85 &2.10\cite{bridge1966polarization} \\
            HBr &2.43 &2.45 &1.98 &1.92 &1.65 &2.35 &2.30 &2.02 &1.7\cite{pinkham2008extracting}\\
            HI  &2.66 &2.81 &2.09 &2.00 &1.87 &2.60 &2.51  & &11.4\cite{denbigh1940polarisabilities}\\
            ICl &26.14 &26.96 &24.27 &24.63 &24.50 &25.30 &25.82 & & \\
            I$_{2}$ &44.92 &49.01  &41.31 &43.75 &42.88 &44.00 &45.87 & &45.1$\pm$ 2.3\cite{maroulis1997electrooptical}\\
            NaLi &92 &92 &109 &109 &108 &154 &154 &149 &\\
            CsK &353  &362 &400 &389 &390 &510\tnote{d} &499\tnote{d} & &\\
            \hline
            \end{tabular}}
            \begin{tablenotes}
                \item [a] Nonrelativistic calculation using the Levy-Leblond Hamiltonian
                \item [b] Relativistic calculation using the X2C Hamiltonian
                \item [c] Using diffuse Triple-zeta basis set
                \item [d] Correlate both 6s and 5p electrons of Cs
            \end{tablenotes}
        \end{threeparttable}
        \end{table}

    We now turn our attention to the frequency-dependent polarizability and focus on the I$_{2}$ molecule given the extensive experimental research on this molecule and the abundance of experimental spectral data that can be used to validate theoretical models. Relevant in the frequency region that we consider are the lowest electronically excited states, that arise from the $\sigma_g^2\pi_u^4\pi_g^3\sigma_u^1$ and $\sigma_g^1\pi_u^4\pi_g^4\sigma_u^1$ configurations. These are primarily triplet states that are denoted in the literature\cite{NISTdiatomic} as $A^3\Pi_u$, $B^3\Pi_u$ and $C^3\Sigma_u$, with the latter originating from the second configuration. The lowest singlet state is from the first configuration and is indicated as $^1B^{''}$. In Table \ref{tab:fre-dep-pol of i2}, we present the computed frequency-dependent polarizability for three theoretical methods alongside the experimental values\cite{maroulis1997electrooptical} measured by \citeauthor{maroulis1997electrooptical} at three frequencies. Like the experimental values, the values computed with CC at these three frequencies are quite close to each other and we find reasonable agreement with the CC values slightly underestimating the experimental data. The HF and B3LYP results deviate rather strongly from the experimental results for the first two frequencies which can be rationalized as being caused by an error in the position of the pole close to the first two laser frequencies, that is computed at a too low energy with HF and DFT(B3LYP). Due to the selection rules for this transition to the B$^3\Pi_{0^+u}$ state, this then leads to a negative value of the parallel (zz-)component of the polarizability for HF and B3LYP, while the perpendicular (xx-)component is not affected and has a similar value for HF, B3LYP and CC.

    \begin{table}[H]
        \centering
        \caption{Frequency dependent polarizability (a.u.) of I$_{2}$ molecule}
        \setlength{\tabcolsep}{12.0 mm}{
        \begin{tabular}{cccc}
             &$\alpha_{zz}(\omega)$  &$\alpha_{xx}(\omega)$ &$\alpha(\omega)$ \\
        \hline
            \multicolumn{4}{c}{$\omega_{1}$=15798 cm$^{-1}$}\\
        \hline
        HF&152.0  &55.0  &87.4  \\
        B3LYP&$-$10.7  &58.7  &35.6  \\
        CC&114.8  &55.8  &75.5  \\
        Exp\cite{maroulis1997electrooptical}&  &  &86.8$\pm$2.2  \\
        \hline
            \multicolumn{4}{c}{$\omega_{1}$=16832.3 cm$^{-1}$}\\
        \hline
        HF&$-$97.3  &56.0  &4.9  \\
        B3LYP&75.4  &62.0  &66.5  \\
        CC&124.0  &56.8  &79.2  \\
        Exp\cite{maroulis1997electrooptical}&  &  & 93.6$\pm$3.4  \\
        \hline
            \multicolumn{4}{c}{$\omega_{1}$=30756.9 cm$^{-1}$}\\
        \hline
        HF&117.9  &55.3  &  76.2\\
        B3LYP&114.5  &61.0  &78.8  \\
        CC&113.9  &59.9  &77.9  \\
        Exp\cite{maroulis1997electrooptical}&  &  &95.3$\pm$1.9  \\
        \hline
        \end{tabular}}
        
        \label{tab:fre-dep-pol of i2}
    \end{table}

\begin{figure}[H]
    \centering
    \includegraphics[width=15cm,height=11cm]{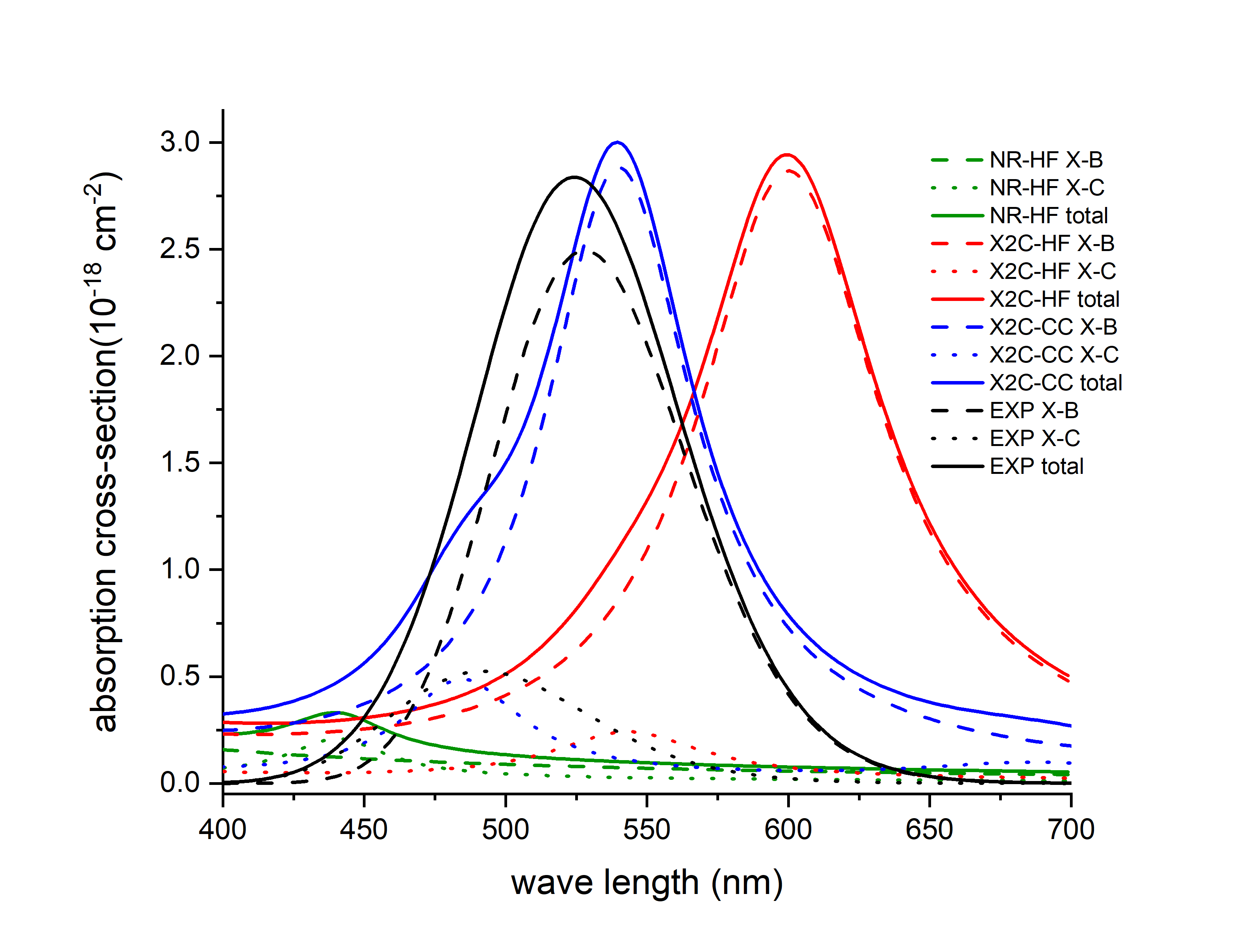}
    \caption{Simulated Spectrum of I$_{2}$. The green lines are non-relativistic HF, the red lines are X2C-HF, the blue lines are X2C-CC, and the black lines are experimental value from work of \citeauthor{tellinghuisen2011least}\cite{tellinghuisen2011least}. The dashed lines represent the contribution of transition from ground state X$^{1}\Sigma_{0+}$ to the B$^{3}\Pi_{0+}$ state. The dot lines are contributions of transition from X$^{1}\Sigma_{0+}$ state to C$^{1}\Pi_{1}$ state. The solid lines are total absorption cross-section.}
    \label{fig:i2_cpp}
\end{figure}

    Rather than looking at the values for just these three frequencies, two of which are close to the X$\rightarrow$B transition, it is more illustrative to apply 
    Eq.~\eqref{eq:abs_cpp} and plot simulated absorption cross-section curves. We scan the wavelength ranging from 400 nm to 700 nm and set the imaginary component of the complex frequency ($\gamma$) to 0.005 a.u, which corresponds to the experimental lifetime of the B$^{3}\Pi_{0+}$ state. As selection rules are different for the transitions to the B$^{3}\Pi_{0+}$ and C$^{1}\Pi_{1}$ states we may thereby identify the $zz$-component of the complex dipole electric polarizability as being (primarily) due to the B$^{3}\Pi_{0+}$ state, while the $xx$-component is due to the C$^{1}\Pi_{1}$ state. This facilitates the comparison to the experimental analysis that was carried out by \citeauthor{tellinghuisen2011least}\cite{tellinghuisen2011least}. The resulting curves for three models, NR-HF(green lines), X2C-HF(red lines), and X2C-CC(blue lines), are depicted in Fig.~\ref{fig:i2_cpp} and clearly show the effect of SOC. The NR computed curves are entirely due to the weaker transition X$^{1}\Sigma_{0+}$ to C$^{1}\Pi_{1}$ and severely underestimate the absorption cross-section. With SOC, this transition becomes a shoulder on the dominant X$^{1}\Sigma_{0+}$ to B$^{3}\Pi_{0+}$ transition. Comparison with the measured curves (black lines) from the work of \citeauthor{tellinghuisen2011least}\cite{tellinghuisen2011least} shows a quite good agreement for the height of the dominant peak that is slightly red-shifted compared to the experimental transition.

    The dominant peak in the X2C-HF exhibits a severe red-shift, which clarifies the error seen in the frequency-dependent polarizability in Table \ref{tab:fre-dep-pol of i2}. For the peak values, the X2C-HF result 2.95 $10^{-18}$cm$^{-2}$ is, however, quite close to the X2C-CC value 2.98 $10^{-18}$cm$^{-2}$ suggesting that the value of the transition dipole moment is similar in both models under the current calculation conditions. This suggest that the relativistic HF model does describe this excited state qualitatively well, albeit at a wrong energy.
    
    Regarding the spin-allowed transition from X$^{1}\Sigma_{0+}$ to C$^{1}\Pi_{1}$ state, displayed by the dotted line in Fig \ref{fig:i2_cpp}, we observe the X2C-CC model to agree well with the experimental analysis of \citeauthor{tellinghuisen2011least}\cite{tellinghuisen2011least}. The discrepancy in pole location is around 20 nm and the difference in the peak value is minor at 0.02 $10^{-18}$cm$^{-2}$. 

    To further analyze the results, we calculate and present the excitation energy of B$^{3}\Pi_{0+}$ and C$^{1}\Pi_{1}$ states in Table \ref{tab:i2_exc}. We note that HF underestimates the excitation energy of B$^{3}\Pi_{0+}$ compared to the CC values no matter what relativistic effects are included. This can be attributed to the triplet instability of the TDHF model\cite{dreuw_single-reference_2005,seeger_self-consistent_1977,bauernschmitt_stability_1996,sears_communication_2011,peach_influence_2011,peach_triplet_2013,lutnaes_spinspin_2010,foresman_toward_1992,rishi2020route}. On the other hand, such underestimation can be largely avoided by using the simpler configuration interaction singles (CIS) approach. Thus, we perform nonrelativistic CIS calculations by DALTON and find that the CIS value indeed is higher by about 0.011 a.u. for the B$^{3}\Pi_{0+}$ state.
    
    For the singlet state C$^{1}\Pi_{1}$, HF excitation energies are larger than the CC values in both nonrelativistic and spin-free calculations. However, when spin-orbit coupling is introduced via X2C, HF energies become lower. While SOC raises the excitation energy in CC, it reduces the excitation energy in HF, implying that SOC effects and correlation are again not additive.

               \begin{table}[H]
        \begin{threeparttable}
            
            \center
            \caption{Excitation energy (a.u.) for I$_{2}$}\label{tab:i2_exc}

        \setlength{\tabcolsep}{3.5mm}{
            \begin{tabular}{cccccccc}
            States&  NR-HF &NR-CIS\tnote{a} &NR-CC &SF-HF &SF-CC& X2C-HF& X2C-CC\\
            \hline
            B$^{3}\Pi_{0+}$    &0.065 &0.076 &0.076 &0.060 &0.070 &0.076 &0.084\\
            C$^{1}\Pi_{1}$     &0.103 &0.107 &0.098 &0.097 &0.092 &0.084 &0.094\\
            \hline
            
            \end{tabular}}

            \begin{tablenotes}
                \item[a] Calculations were performed using the DALTON program
            \end{tablenotes}

        \end{threeparttable}
        \end{table}

In the supplemental materials, we provide a calculation of the spectrum of BH molecule which was used to verify the correctness of our implementation. We compare our damped CC-LR calculation with the broadening of coupled cluster transition dipole moment computed by the DALTON program\cite{aidas2014d,christiansen1998integral}. and find good agreement.

\subsection{Spin-spin coupling }

In the previous section, we investigated electric properties. In the current section, we show a calculation of the indirect nuclear spin-spin coupling constant as an illustrative example of the use of our implementation for a magnetic property. The coupling constant $K_{KL}$ can be related to the experimentally observed coupling $J_{KL}$ between the nuclear spins of atoms $K$ and $L$ via
\begin{equation}
    J_{KL} = \frac{1}{2\pi}\gamma_{K}\gamma_{L}K_{KL}
\end{equation}

\noindent where $\gamma_{K}$ is the gyromagnetic ratio of nucleus $K$.
The $K_{KL}$ tensor can in a relativistic framework be expressed in terms of linear response functions with respect to the hyperfine operator $\hat{h}_{K}^{\mathrm{hfs}}$\cite{helgaker_recent_2012}:
\begin{equation}
    K_{KL,\mu\nu} =  \frac{\partial^{2}}{\partial m_{K;\mu}\partial m_{L;\nu}}\langle\langle \hat{h}_{K}^{\mathrm{hfs}}; \hat{h}_{L}^{\mathrm{hfs}} \rangle\rangle_{\omega_{k1}, \omega_{k2}}
    \label{spsp_rel_lr}
\end{equation}
\begin{equation}
\label{h_mag_4C}
    \hat{h}_{K}^{\mathrm{hfs}} =  \frac{1}{c}\sum_{i}\textbf{m}_{K}\cdot\frac{\textbf{r}_{iK}\times  \pmb{\alpha}}{r_{iK}^{3}}
\end{equation}

In the non-relativistic framework, it is common to formulate $K$ in terms of three distinct 
contributions: diamagnetic spin-orbit coupling (DSO), paramagnetic spin-orbit (PSO), and the Fermi contact-Spin dipolar (FC-SD) term. Of these, the first term can be computed as an expectation value, whereas the second and third require the use of response theory. Moreover, the PSO term involves only singlet excitations, whereas the FC-SD term couples a singlet ground state to triplet excited states due to the triplet nature of the Fermi contact and spin-dipolar operators. An explicit sum-over-states form of the contributions to $K_{KL}$ in the nonrelativistic framework is\cite{helgaker_recent_2012,ramsey1953electron}:

\begin{equation}
\begin{split}
    K_{KL} &= \frac{1}{2c^4}
    \bra{0}\frac{\textbf{r}_{K}^{T}\textbf{r}_{L}I-\textbf{r}_{K}\textbf{r}_{L}^{T}}{r_{K}^{3}r_{L}^{3}}\ket{0}-\frac{2}{c^4}\sum_{n_{S}}\frac{\bra{0}r_{K}^{-3}I_{K}\ket{n_{S}}\bra{n_{S}}r_{L}^{-3}I_{L}\ket{0}}{E_{n_{S}}-E_{0}} \\
    &
    -\frac{2}{c^4}\sum_{n_{T}}\frac{\bra{0}\frac{8\pi}{3}\delta(r_{L})\textbf{s}+\frac{3\textbf{r}_{L}\textbf{r}_{L}^{T}-r_{L}^{2}I_{3}}{r_{L}^{5}}\textbf{s}\ket{n_{T}}\bra{n_{T}}\frac{8\pi}{3}\delta(r_{L})\textbf{s}^{T}+\frac{3\textbf{r}_{L}\textbf{r}_{L}^{T}-r_{L}^{2}I_{3}}{r_{L}^{5}}\textbf{s}^{T}\ket{0}}{E_{n_{T}}-E_{0}}
\end{split}
\label{spsp_nonrel_lr}
\end{equation}

As discussed in Reference~\citenum{aucar1999origin}, the PSO and FC-SD response functions can in the relativistic framework of Eq. \eqref{spsp_rel_lr} be identified as orbital responses involving rotations amongst positive energy orbitals. The DSO contribution, on the other hand, comes from the rotations between positive and negative energy orbitals and can in a sequence of approximations be brought into an expectation value form that is identical to the non-relativistic expression and is then called the Sternheim approximation\cite{sternheim1962second}. 
Therefore, in relativistic calculations, there are two ways to obtain the diamagnetic terms: one by including electron-positron rotations explicitly in the response calculation or by making use of the Sternheim approximation. 

In contrast to the Sternheim approximation, in which a numerically very stable expectation value is computed, the formally more rigorous response approach is quite sensitive to the quality of sampling of the positronic orbital space in a finite basis\cite{aucar1999origin,visscher1999full}. This is why in the current study, we compute the diamagnetic terms as an expectation value. An important modification as compared to the original application in 4-component theory is the use of the X2C transformation, in which all operators are first transformed to a 2C representation. The generic expression is:

\begin{equation}
    \mathbf{X}^{X2C} = \mathbf{X}^{++} = \sum_{V,W}^{L,S} {[\mathbf{U}^{V+}]}^\dagger\mathbf{X}^{VW}\mathbf{U}^{W+}
\end{equation}

in which $\mathbf{U}^{L+}$ and $\mathbf{U}^{S+}$ are blocks of the X2C transformation matrix that block-diagonalizes the matrix representation of a reference Hamiltonian operator 
(usually and also in this work taken as the molecular Hamiltonian without 2-electron interactions) and allows to solve only for positive (+) energy solutions. 
In case of magnetic properties, the original 4C matrix representation of operator (\ref{h_mag_4C}) is off-diagonal with respect to the Large (L) and (S) Small parts of the 4C basis so that

\begin{equation}
    \mathbf{h}_K^{hfs,X2C} = \frac{1}{c} {[\mathbf{U}^{L+}]}^\dagger [\textbf{m}_{K} \cdot \frac{\boldsymbol{\sigma}\times\mathbf{r}_{iK}}{r_{iK}^{3}}]^{LS}
    \mathbf{U}^{S+} + h.c.
\end{equation}

This matrix representation can be interpreted as providing the X2C equivalent of the singlet PSO and triplet FC-SD operators that are used in nonrelativistic response calculations. The Sternheim approximation yields a diagonal 4C DSO operator $\mathbf{k}_{KL}$ that is transformed as 
\begin{eqnarray}
    \mathbf{k}_{KL}^{X2C} = {[\mathbf{U}^{L+}]}^\dagger \mathbf{k}_{KL}^{LL} \mathbf{U}^{L+} + 
    {[\mathbf{U}^{S+}]}^\dagger \mathbf{k}_{KL}^{SS} \mathbf{U}^{S+} \\
    \mathbf{k}_{KL}^{VW} =\frac{1}{2c^4} [\frac{\textbf{r}_{K}^{T}\textbf{r}_{L}I-\textbf{r}_{K}\textbf{r}_{L}^{T}}{r_{K}^{3}r_{L}^{3}}]^{VW}\delta_{VW}
\end{eqnarray}

and contracted with the unperturbed density matrix to obtain the DSO contribution to the spin-spin coupling. 

In Table \ref{tab:spsp}, we list the resulting reduced isotropic and anisotropic spin-spin coupling constants of HX(X=F, Cl, Br, I) computed by HF, B3LYP, CC-CI, and CC-CC models with both nonrelativistic and relativistic Hamiltonians. As is well-known, relativistic effects are very important for magnetic properties and we see the expected increase of their magnitude upon descending 
the periodic table from hydrogen fluoride to hydrogen iodide. To benchmark the quality of the X2C transformation, we also carried out four-component Dirac-Coulomb(DC) HF calculations with default approximation for the all small two-electron integrals\cite{visscher1997approximate} and see that the X2C values match the DC results very well for all molecules.  

At the Hartree-Fock level, the isotropic constants generally exhibit a downward trend from HF to HI, while the anisotropic values typically show an upward trend for both relativistic and nonrelativistic calculations. After including electron correlation, these trends are qualitatively the same although the precise values change considerably, especially for HBr. To verify our CC implementation, we also utilize the CFOUR program\cite{matthews2020coupled} for nonrelativistic CC response and find our CC-CC models with the nonrelativistic Hamiltonian to reproduce the CFOUR values for all three light molecules. 

Although the nonrelativistic calculation is useful for analysis, we cannot ignore relativistic effects for heavy molecules. For example, the relativistic correction at the coupled cluster level for HBr is around 25\% and slightly smaller than that with Hartree-Fock. We also performed DFT calculations, the results obtained with B3LYP functionals are quite far from both the HF and the CC results. As there are no suitable experimental values to compare with one cannot assess rigorously the performance of the methods, but the large discrepancy between the commonly used B3LYP DFT and CC makes these systems of interest for future benchmarking with converged CC expansion (we deem both our employed basis set as well as excitation level not yet suitable for this purpose).

Looking at the two ways of carrying out CC response calculations, we observe minor variances between the CC-CI and CC-CC, which appear to become more pronounced for the heavier elements. It is known that LR-CC transition moments are size-extensive whereas EOM-CC ones are not ~\cite{kobayashi1994calculation,Koch1994,Sekino1994,Sekino1999,caricato2009difference,Perera2010,Nanda2015,coriani_molecular_2016,faber2018resonant}, though in these comparisons it was found the numerical differences between LR-CC and EOM-CC were rather small for a single molecule. Numerical studies have been primarily concerned with light molecules and properties within the valence domain, like the electric transition dipole moment\cite{caricato2009difference,Nanda2015,faber2018resonant}, and the dipole polarizability\cite{kobayashi1994calculation,Sekino1999}, and our results for polarizabilities are in line with these findings. A notable exception is the work of~\citet{Sekino1999}, which have investigated spin-spin couplings for ethane and found a difference of 0.05\% between EOM-CC and LR-CC for $J_{CC}$. This value is comparable to our difference of about 0.19\% for the HF molecule.

If the lack of size extensivity in EOM-CC transition moments is a significant source of discrepancies, one would expect the difference between CC-CI and CC-CC to grow as the number of electrons correlated increases across the HX series, but the difference per correlated electron to remain roughy constant. Our analysis of the $zz, xx$ and $yy$ components of the linear response contribution to $K_{HX}$ (see supplementary information) provides some evidence this is the case, as differences (in absolute value) for each component fall between 0.004 and 0.02 a.u.\ for all molecules. There are some differences between Hamiltonians for HBr and HI, but these are of smaller magnitude than those due to non-extensivity. However, we believe the sample size is not large enough for definitive conclusions, and in future investigations we intent to revisit this issue for a broader range of molecules.

               \begin{table}[H]
        \begin{threeparttable}
            
            \center
            \caption{Isotropic and anisotropic reduced spin-spin coupling K(10$ ^{19}$ m$^{-2}$ kg s$^{-2}$ A$^{-2}$) for HX(X=F, Cl, Br, I)}\label{tab:spsp}

        \setlength{\tabcolsep}{8.0mm}{
            \begin{tabular}{ccccc}
            Models&  $^{1}$HF$^{19}$ &$^{1}$HCl$^{35}$ &$^{1}$HBr$^{79}$ &$^{1}$HI$^{127}$ \\
            \hline
             \multicolumn{5}{c}{Isotropic}\\
            \hline
            NR-HF     &49.5486 &28.1528 &10.8253 &-0.8979\\
            NR-B3LYP  &33.3898 &19.7146 &-1.8769 &\\
            NR-CC-CI  &40.5554 &31.3181 &30.7926 &\\
            NR-CC-CC  &40.4794 &31.0971 &29.9730 &\\
            NR-CC-CC\tnote{a}     &40.4778 &31.0970 &29.9729 &\\
            X2C-HF    &49.5023 &27.2261 &-4.5338 &-83.1522\\
            X2C-B3LYP &33.2367 &18.9409 &-11.6914 &-57.3316\\
            X2C-CC-CI &40.4834 &30.9008 &23.8246 &3.4887\\
            X2C-CC-CC &40.4047 &30.6448 &22.7588 &0.7481\\
            DC-HF     &49.4725 &27.1494 &-4.8396 &-84.0079\\
            \hline
            \multicolumn{5}{c}{Anisotropic}\\
            \hline
            NR-HF     &2.5499 &59.6666 &161.9806 &277.7237\\
            NR-B3LYP  &6.3484 &50.1075 &130.4249 &\\
            NR-CC-CI  &-3.7566 &36.3828 &100.9785 &\\
            NR-CC-CC  &-3.4931 &37.1193 &102.9362 &\\
            X2C-HF    &2.5858 &60.2375 &168.5425 &305.7204\\
            X2C-B3LYP &6.4477 &50.3990 &130.5655 &201.0597\\
            X2C-CC-CI &-3.6579 &36.8838 &106.6559 &192.8454\\
            X2C-CC-CC &-3.3929 &37.6281 &108.7226 &196.3474\\
            DC-HF     &2.5978 &60.2822 &168.7214 &306.1280\\
            \hline
            
            \end{tabular}}

            \begin{tablenotes}
                \item[a] Calculations were performed using the CFOUR program
            \end{tablenotes}

        \end{threeparttable}
        \end{table}

\begin{figure}[H]
            \centering
            \includegraphics[scale=0.2]{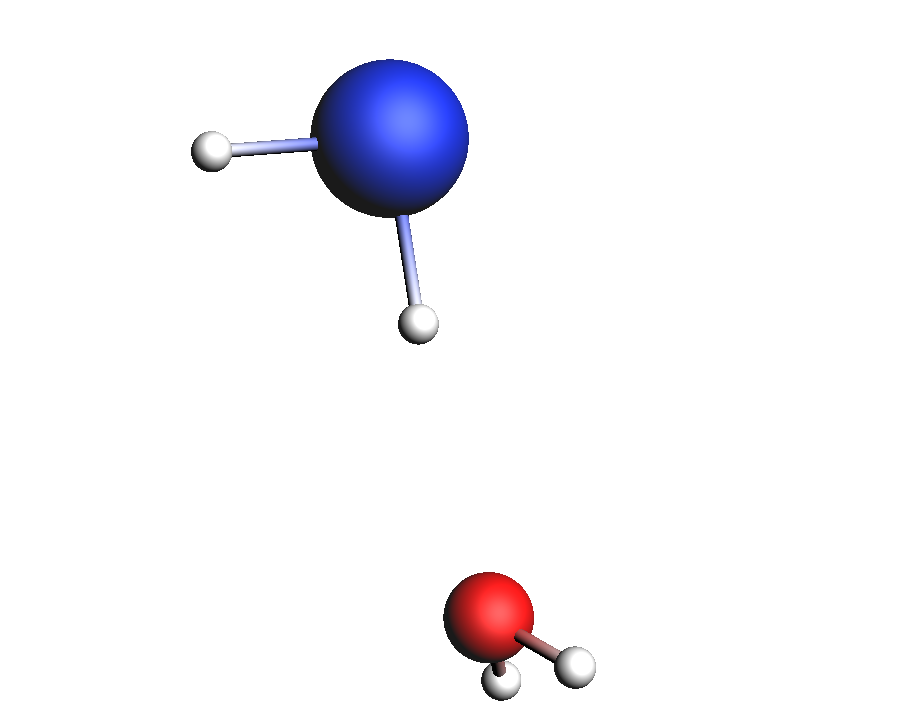}
            \caption{H$_{2}$Se-H$_{2}$O complex system. Color of atoms: Se (blue), O(red), H(white), H$_{b}$ is the Hydrogen atom that belongs to Se and is close to O.}
            \label{fig:h2se-h2o}
        \end{figure}

As most experimental work is carried out in the condensed phase, we wanted to go beyond isolated diatomic molecules, and provide a sample investigatation of solvent effects. For this purpose we chose the solvent shift on the spin-spin coupling constant $^{1}$H$_{b}$-${^{34}}$Se in the the H$_2$Se-H$_2$O dimer. The supermolecular structure is taken from the work of \citeauthor{olejniczak2017calculation}\cite{olejniczak2017calculation} and displayed on Fig.\ref{fig:h2se-h2o}. It can readily be seen from Table \ref{tab:solvent of h2se} that all calculations show the solvent effect on the Se-H$_{b}$ coupling for the bond involved in the hydrogen bond to be quite substantial. However, the shifts $\Delta J$ in the correlated models have a different magnitude than that at the HF level. For example, the shifts of Se-H$_{b}$ are 19.5403 Hz and 19.0648 Hz for CC-CC and B3LYP, respectively, while they are almost twice as large at 40.0220 Hz for HF. Although, the shifts of DFT are quite close to those computed with CC, the absolute $J_{iso}^{super}$ and $J_{iso}$ deviate a lot. Comparing with BLYP and B3LYP values, we find the addition of exact exchange to the DFT to have a significant effect, with the hybrid DFT B3LYP results being closer to the CC values.

    \begin{table}[H]
    \begin{threeparttable}

        \centering
        \caption{Isotropic and anisotropic indirect spin-spin coupling ($J_{iso}$ and $J_{aniso}$ in Hz) for isolated H$_{2}$Se subsystem, ($J_{iso}^{super}$ and $J_{aniso}^{super}$ in Hz) for H$_{2}$Se subsystem in H$_{2}$Se-H$_{2}$O, and the shifts ($\Delta J$, in Hz) for the isolated ("ME") H$_{2}$Se molecules in the presence of H$_{2}$O}
        \setlength{\tabcolsep}{5.0mm}{
        \begin{tabular}{ccccccc}
        Models  & $J_{iso}^{}$ & $J_{iso}^{super}$ & $\Delta J_{iso}^{ME}$  & $J_{aniso}^{}$ &$J_{aniso}^{super}$ &$\Delta J_{ianso}^{ME}$ \\
        \hline
            \multicolumn{7}{c}{$^{1}$H$_{b}$-Se$^{34}$}\\
        \hline
        HF\tnote{a} &90.4949 &128.2837   &36.7888  &305.8746 &302.1583  &-3.7163 \\
        HF   & 52.7369 & 92.7589 &40.0220  &353.6191 &353.4049  &-0.2142 \\
        BLYP &-26.8400 &-11.0675 &15.7725  &265.5061 &271.8454  &6.3393\\
        B3LYP& -9.1404 & 9.9244  &19.0648  &269.7740 &275.2143  &5.4403 \\
        CC-CI& 66.6432 &85.8755  &19.2303  &215.4586 &218.7408  &3.2821 \\
        CC-CC& 65.9553 &85.4956  &19.5403  &219.8036 &223.0023  &3.1987 \\
        \hline
        \end{tabular}}

        \begin{tablenotes}
            \item[a]  Nonrelativistic calculation with the Levy-Leblond Hamiltonian
        \end{tablenotes}
        \label{tab:solvent of h2se}
        \end{threeparttable}
    \end{table}

\subsection{Optical rotation}

Finally, we consider both electric and magnetic fields, by looking at optical rotation (in the length gauge and for a common gauge origin) for the archetypical chiral molecules H$_{2}$Y$_{2}$(Y=O, S, Se, Te). At the frequency of the sodium D-line (in 589.29 nm), which is the most common experimental setup, the specific optical rotation $[\alpha]_{D}^{25}$ in unit [${^\circ}$ dm$^{-1}$(g/mol)$^{-1}$] is given by the equations\cite{Ruud2002}  

    \begin{equation}
        [\alpha]_{D}^{25}=-228\cdot10^{-30}\frac{\pi^{2}N a_{0}^{4}\omega}{3M}\sum_{\alpha}G'_{\alpha\alpha}
    \end{equation}

    \begin{equation}
        G'_{\alpha\beta}(-\omega;\omega)= -\mathrm{Im}\langle\langle \hat{\mu}_{\alpha}; \hat{m}_{\beta} \rangle\rangle_{\omega}
    \end{equation}

\noindent where $M$ is the molecular mass in g mol$^{-1}$, $N$ is the number density, and $\mu_{\alpha}$ and $m_{\beta}$ are the electric and magnetic dipole operator, respectively. 

\begin{figure}[H]
            \centering
            \includegraphics[width=17cm,height=12cm]{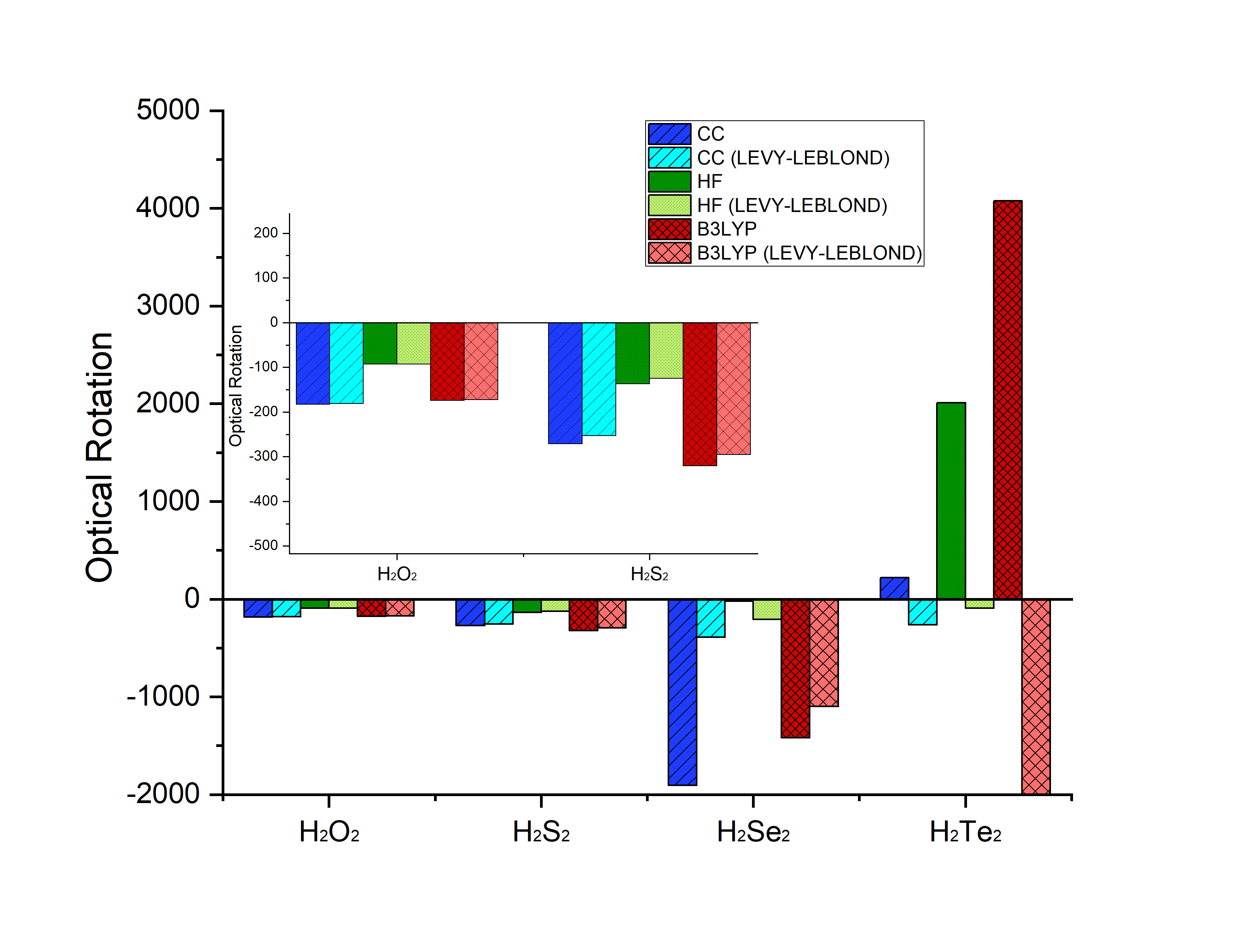}
            \caption{Optical rotation of Hydrogen peroxide series (H$_{2}$Y$_{2}$) in [${^\circ}$ dm$^{-1}$(g/mol)$^{-1}$] with a frequency corresponding to the sodium D-line (589.29 nm, 0.077319 a.u.) calculated with the X2C and Levy-Leblond Hamiltonians}
            \label{fig:opt_rot}
        \end{figure}

In Fig \ref{fig:opt_rot}, we display the results for HF, B3LYP, and CC for both the nonrelativistic and X2C Hamiltonian. First, to verify our implementation, we performed the calculation on H$_{2}$S$_{2}$  with the DALTON program with the same basis set. The resulting data are available in the supplemental information and show good agreement, confirming the correctness of the implementation. To benchmark the influence of the truncation on the virtual orbital space, we furthermore performed a calculation in which we truncated the virtual orbital space with an energy threshold of 100 a.u. instead of the otherwise used value on 5 a.u. and found that results match up to 99\%. This is similar to the tendency observed in the electric dipole polarizability, as expected as both optical rotation and electric dipole polarizability are predominantly determined by the valence electrons and do not require core-like high-energy virtuals. 

Fig ~\ref{fig:opt_rot} shows that for the lighter molecules, H$_{2}$O$_{2}$ and H$_{2}$S$_{2}$, the B3LYP and CC values are nearly twice as large than those of the HF. While the relativistic effect is negligible for H$_{2}$O$_{2}$, with a correction of less than 1\%, it cannot be neglected for H$_{2}$S$_{2}$, where it rises to 10\%. The impact of the relativistic effect is present for all models, but correlation and relativistic effects are again not additive. For instance, we find a relativistic HF correction of -12 [${^\circ}$ dm$^{-1}$(g/mol)$^{-1}$], while for B3LYP and CC, these corrections are -26 [${^\circ}$ dm$^{-1}$(g/mol)$^{-1}$] and -18 [${^\circ}$ dm$^{-1}$(g/mol)$^{-1}$], respectively.
For the heavier molecules H$_{2}$Se$_{2}$ and H$_{2}$Te$_{2}$, values computed for the sodium D-line frequency become exceedingly large as these molecules have an excitation that is almost at resonance with this frequency. To better understand this phenomenon, we have therefore calculated the excitation energy of the first eleven microstates for these two molecules. The resulting values are compiled and presented in Table \ref{tab: H2Y2exc}. Note that we display all degenerate components of triplet states for better comparison to relativistic states.

The computed excitation values show that in case of H$_{2}$Se$_{2}$, the relativistic CC value is significantly larger than the nonrelativistic CC because the employed frequency is quite close to the resonance frequency of the second excited state in the relativistic calculation (0.0789 a.u.), whereas it is distant from all excited states in the nonrelativistic CC calculation. For the B3LYP computations, we see that the frequency is close to the fourth excited state in both relativistic and nonrelativistic scenarios (0.0863 and 0.0889 respectively). This proximity results in large values being obtained from both calculations.

When we examine the H$_{2}$Te$_{2}$ molecule, we find the relativistic effect to be substantial for all three models and even reversing the sign of the optical rotation. For example, the nonrelativistic CC value is -263.59 [${^\circ}$ dm$^{-1}$(g/mol)$^{-1}$], but the relativistic CC is 218.83 [${^\circ}$ dm$^{-1}$(g/mol)$^{-1}$]. Besides reversing the sign, with HF also the magnitude of the optical rotation is very different in the relativistic and nonrelativistic cases. This is because the first six excited states, while being close to transitions, are triplets and hence do not contribute to the optical rotation that is in the nonrelativistic case. In the relativistic case, SOC makes these transitions allowed, which combined with their proximity to the sodium D-line leads to a much larger optical rotation of opposite sign than computed non-relativistically. The B3LYP values are large in both the relativistic and the NR case because the frequency is then close to the singlet state (0.0721 a.u. and 0.0777 a.u. respectively). To avoid artifacts due to the proximity of poles and the associated infinity of the real frequency-dependent response function, it is probably opportune to consider the lifetime of the excited state and use damped response theory like shown for the complex polarizability for I$_{2}$.

In addition, we observe the triplet instability in HF results as well, similar to what we observed for I$_{2}$. For example, we note the excitation energy of HF's first triplet is larger than that in CC. In HF calculations, the second triplet state lies below the first singlet state. However, in both the correlation models B3LYP and CC, the first singlet state is positioned above the second triplet. To address this issue, we perform nonrelativistic calculations for ten lowest states including five triplets and five singlets. Detailed results are provided in the supplementary information. Similar to I$_{2}$, we note the CIS largely overcomes the triplet instability seen in HF and yields a more systematic error pattern when compared to B3LYP and CC.

        \begin{table}[H]
        \begin{threeparttable}

            \center
            \caption{Excitation energy (a.u.) of the first eleven microstates for H$_{2}$Se$_{2}$ and H$_{2}$Te$_{2}$}.\label{tab: H2Y2exc}

        \setlength{\tabcolsep}{3.2mm}{
            \begin{tabular}{ccccccccc}
            &State& HF &HF\tnote{a} & B3LYP &B3LYP\tnote{a} &CC &CC\tnote{b} & CC\tnote{a}\\
            \hline
            H$_{2}$Se$_{2}$&1 &0.0671&0.0719 &0.0639 &0.0661 &0.0789 &0.0785 &0.0812 \\
                           &2 &0.0672&0.0719 &0.0639 &0.0661 &0.0789 &0.0785 &0.0812 \\
                           &3 &0.0699&0.0719 &0.0642 &0.0661 &0.0792 &0.0788 &0.0812 \\
                           &4 &0.0860&0.0954 &0.0863 &0.0889 &0.0988 &0.0984 &0.1016 \\
                           &5 &0.0911&0.0954 &0.1098 &0.1121 &0.1285 &0.1282 &0.1309 \\
                           &6 &0.0913&0.0954 &0.1099 &0.1121 &0.1286 &0.1282 &0.1309 \\
                           &7 &0.1082&0.1083 &0.1100 &0.1121 &0.1288 &0.1284 &0.1309 \\
                           &8 &0.1289&0.1315 &0.1314 &0.1343 &0.1443 &0.1440 &0.1478 \\
                           &9 &0.1293&0.1315 &0.1497 &0.1510 &0.1541 &0.1541 &0.1561 \\
                           &10&0.1295&0.1315 &0.1499 &0.1510 &0.1543 &0.1542 &0.1561 \\
                           &11&0.1600&0.1624 &0.1500 &0.1510 &0.1545 &0.1544 &0.1561 \\
            \hline
            H$_{2}$Te$_{2}$&1 &0.0517 &0.0635 &0.0542 &0.0584 &0.0672 &0.0672 &0.0721 \\
                           &2 &0.0517 &0.0635 &0.0543 &0.0584 &0.0673 &0.0673 &0.0721 \\
                           &3 &0.0605 &0.0635 &0.0558 &0.0584 &0.0689 &0.0686 &0.0721 \\
                           &4 &0.0697 &0.0924 &0.0721 &0.0777 &0.0835 &0.0835 &0.0898 \\
                           &5 &0.0873 &0.0924 &0.0931 &0.0968 &0.1098 &0.1097 &0.1143 \\
                           &6 &0.0888 &0.0924 &0.0935 &0.0968 &0.1101 &0.1100 &0.1143 \\
                           &7 &0.0993 &0.0957 &0.0940 &0.0968 &0.1109 &0.1108 &0.1143 \\
                           &8 &0.1098 &0.1138 &0.1102 &0.1158 &0.1220 &0.1220 &0.1292 \\
                           &9 &0.1123 &0.1138 &0.1316 &0.1320 &0.1349 &0.1349 &0.1367 \\
                           &10&0.1139 &0.1138 &0.1324 &0.1320 &0.1357 &0.1356 &0.1367 \\
                           &11&0.1370 &0.1402 &0.1324 &0.1320 &0.1360 &0.1359 &0.1367 \\
            \hline
            \end{tabular}}
                \begin{tablenotes}
                    \item[a]  Nonrelativistic calculation with the Levy-Leblond Hamiltonian
                    \item[b] With trucation of virtual orbitals on 100 a.u. donce by RELCCSD.
                \end{tablenotes}
            \end{threeparttable}

        \end{table}

\section{Conclusion}
\label{Conclusions}
In this work, we describe the formulation and implementation of the relativistic coupled cluster linear response method for static and frequency-dependent molecular property calculations, which can accurately treat both relativistic and electronic correlation effects. 
This implementation was accomplished in the GPU-accelerated coupled cluster module of the DIRAC program leveraging a framework designed to handle similar transformed Hamiltonian in subspace. This framework aids in solving both eigenvalue and linear system problems. The current code is capable of calculating excitation energies within the EOM-CCSD framework and computing the linear response function for both CC-CI and CC-CC type wave-function models. 

We have validated the implementation by assessing purely electric properties such as static and frequency-dependent polarizability for Group IIB atoms (Zn, Cd, Hg) and several diatomic molecules. Compared to 
previous Hartree-Fock linear response calculations, our current linear response calculation based on the relativistic coupled cluster approach offers  a notably improved accuracy. This enhancement is particularly evident in terms of relativistic corrections and correlation, bringing our results  closer to the experimental data. 

In this study, we also tested the indirect spin-spin coupling constant---a purely magnetic property--- for the hydrogen halide series HX(X=F, Cl, Br, I). Validation was done by reproducing the results obtained by other programs such as DALTON and CFOUR using a nonrelativistic Hamiltonian. We extended our study to explore the impact of solvent effect on the H$_{2}$Se-H$_{2}$O complex systems. Both correlation and relativistic corrections were found to have pronounced effects on the solvent shift. While CC and DFT gave  similar magnitudes for the shifts in solvent effect, the absolute spin-spin coupling constants differed significantly. This finding calls for caution when employing DFT for such calculations.

Lastly, we computed the optical rotation--- an electric and magnetic mixed property--- for chiral molecules H$_{2}$Y$_{2}$ (Y=O, S, Se, Te) at the wave-length of sodium D-line (589.29 nm). Our exploration revealed potential challenges when using this frequency for heavy molecules. We analyzed the poles of the response function by calculating the excitation energy, and advise caution when using sodium D-line for these heavier molecules in future investigations. 

A distinguishing aspect of our implementation is its use of complex algebra, which facilitates a straightforward extension of real to complex frequencies for the evaluation of the damped linear response function. We used this feature to simulate the spectrum of I$_{2}$ through the assessment of the absorption cross-section.

As a final point and perspective,  it is worth noting that our current implementation relies on the single-code tensor operation library TAL-SH. While efficient, this library is limited to using the memory capacity of a single node. Therefore, a natural 
development
is to extend the current code for the EOM-CCSD energy and linear response to use a library suited for distributed memory computing architectures, such as the ExaTENSOR library already employed for the CC energy evaluation, but still lacks some features needed in the Davidson diagonalization procedure. After resolving these issues we are optimistic that we can eliminate the limitations caused by the library being able to use the memory of only a single compute node  and enable treatment of larger systems.

\begin{acknowledgement}

This research used resources of the Oak Ridge Leadership Computing Facility, which is a DOE Office of Science User Facility supported under Contract DE-AC05-00OR22725 (allocations CHM160, CHM191 and CHP109). XY, LH and ASPG acknowledge funding from projects Labex CaPPA (Grant No. ANR-11-LABX-0005-01) and CompRIXS (Grant Nos. ANR-19CE29-0019 and DFG JA 2329/6-1), the I-SITE ULNE project OVERSEE and MESONM International Associated Laboratory (LAI) (Grant No. ANR-16-IDEX-0004), as well support from the French national supercomputing facilities (Grant Nos. DARI A0130801859, A0110801859, project grand challenge AdAstra GDA2210).
SC acknowledges funding from the Independent Research Fund Denmark--Natural Sciences, Research Project 2 - grant no. 7014-00258B.

\end{acknowledgement}


\begin{suppinfo}

The data (input/output) corresponding to the calculations of this paper are available at the Zenodo repository under DOI: \href{http://doi.org/10.5281/zenodo.8136133}{10.5281/zenodo.8136133}.

\end{suppinfo}

\bibliography{ms}

\end{document}






\section{Working equations for CCSD linear response}

In what follows $a,b,c,.. $ will indicate {particle lines}, $ i,j,k,... $ {hole lines}, and $ p,q,r,s,... $ general indexes~\cite{crawford2007introduction}. In all equations below we use Einstein notation. Furthermore we define 
\begin{itemize}
    \item $P$ as a permutation operator, with : $P_{-pq} f\left(\dots pq \dots\right)= f\left(\dots pq \dots\right) - f\left(\dots qp \dots\right)$;
    \item $X_{q}^{p} = \bra{p} X \ket{q}$ are matrix elements of property operator $X$ ;
    \item $V_{rs}^{pq} = \bra{pq}\ket{rs}$ are antisymmetrized two-electron integrals, and $f_{q}^{p} = \bra{p} f \ket{q}$ Fock matrix elements;
    \item $\boldsymbol{\lambda}$ denotes ground-state CC Lagrange multipliers, and is therefore equivalent to $\bar{\mathbf{t}}^{(0)}$, and we have $\bra{\Lambda} = \bra{R} + \sum_\mu \lambda_\mu \bra{{\mu}}e^{-T_0} = \bra{R} + \sum_\mu \lambda_\mu \bra{\bar{\mu}} \equiv \bra{R} + \sum_\mu \bar{t}^{(0)}_\mu \bra{\bar{\mu}}$
    \item $\mathbf{r}$ and $\mathbf{l}$ denote, depending on context, (trial) vectors associated to the solution of right and left-hand EOMCC or response equations.
\end{itemize}

\subsection{Linear response}


\noindent The $\boldsymbol{\xi}^{X}$ vector is defined as\cite{Christiansen1998}
\begin{align}
\xi^{X}_{\mu} = \bra{\bar{\mu}}X\ketCC \label{xi-equation}
\end{align}
and the programmable expressions for its elements are given by:
\begin{align}
    \xi^{X}{}_{i}^{a} &= + X_{i}^{a} + X_{e}^{a}t_{i}^{e}- X_{i}^{m}t_{m}^{a} - (X_{e}^{m}t_{i}^{e})t_{m}^{a} + X_{e}^{m}t_{im}^{ae}  \\
    \xi^{X}{}_{ij}^{ab}&= + P_{-ab}X_{f}^{b}t_{ij}^{af} - P_{-ij}X_{j}^{m}t_{im}^{ab}-P_{-ij}(X_{e}^{m}t_{i}^{e})t_{mj}^{ab} - P_{-ab}(X_{f}^{m}t_{m}^{a})t_{ij}^{fb}
\end{align}


\noindent The $\boldsymbol{\eta}^{X}$ vector is defined as~\cite{Christiansen1998}
\begin{align}
\eta_\mu^{X}&=\bra{\Lambda}[X,\hat{\tau}_{\mu}]\ketCC \label{eta-equation}
\end{align}
and the programmable expressions for its elements are given by
\begin{align}
	\eta^{X}{}^{i}_{a}=&X^{i}_{a}+X^{e}_{a} \lambda^{i}_{e}- X^{i}_{m}\lambda^{m}_{a}- X^{m}_{a}t^{e}_{m}\lambda^{i}_{e}-X^{i}_{e}t^{e}_{m}\lambda^{m}_{a} -\frac{1}{2}\left(t^{fe}_{mn}\lambda^{mi}_{fe}\right)X^{n}_{a} \nonumber \\
	&-\frac{1}{2}\left(t^{fe}_{nm}\lambda^{nm}_{fa}\right)X^{i}_{e}\\
	\eta^{X}{}^{ij}_{ab} = & +P_{-ij}P_{-ab}\lambda^{i}_{a}X^{j}_{b}+P_{-ab} \lambda^{ij}_{ae} X^{e}_{b}-P_{-i j} \lambda^{i m}_{a b} X^{j}_{m}-P_{-ab} \left(t^{e}_{m} \lambda^{ij}_{ae} \right)  X^{m}_{b} \nonumber \\
	&-P_{-ij} \left(t^{e}_{m} \lambda^{im}_{ab} \right)X^{j}_{e}
\end{align}
whereas the elements of the $^{\textrm{EOM}}\boldsymbol{\eta}$ vector~\cite{faber2018resonant} are
\begin{align}
	{}^{\textrm{EOM}}\eta^{X}{}_{a}^{i} &= +\eta^{X}{}_{a}^{i} +  \lambda^{ij}_{ab}\xi^{X}{}^{b}_{j} 
\end{align}


\noindent The CC Hessian ($\mathbf{F}$) is defined as\cite{Christiansen1998}
\begin{align}
  F_{\mu \nu}=&\left< \Lambda \left | \left[ \left[ H_0,\hat{\tau}_{\mu} \right],\hat{\tau}_{\nu} \right] \right|\mathrm{CC}\right>
\end{align}
and the programmable expressions for the matrix elements for its contraction with a response vector $\tXs{}{}$, $\left( \tXs {}{} F \right) $ are given by
 \begin{align}
 \displaybreak[1]
\left( \tXs {}{} F \right)^{k}_{c} = & +    \V kica . \tXs ai  -\lambdas ic  \F ka . \tXs ai  -\lambdas ka  \F ic . \tXs ai  + \lambdas ke   \W eica . \tXs ai  
- \lambdas mc \W kima . \tXs ai -  \lambdas ma  \W ikmc . \tXs ai + \lambdas ie \W ekac . \tXs ai  \nonumber\\
&
-  \undemi \lambdas ic \V kjab . \tXd abij   - \undemi \lambdas ka   \V ijcb . \tXd abij + \lambdas ia  \V jkbc . \tXd abij   -  \lambdad mkae    \W iemc . \tXs ai - \lambdad imec  \W ekam .  \tXs ai  + \undemi \lambdad kief \W efca . \tXs ai  \nonumber \\
&+ \undemi \lambdad mnca  \W kimn . \tXs ai 
  - \undemi  \left( \lambdad mnec \td efmn \right) \V kifa   \tXs ai  - \undemi \left(\lambdad mkef \td efmn \right) \V nica    . \tXs ai  - \undemi  \left(\lambdad mnea \td efmn \right)  \V ikfc    \tXs ai \nonumber \\
& - \undemi \left(\lambdad mief \td efmn \right)  \V nkac  . \tXs ai  
- \undemi \lambdad ijac \F kb . \tXd abij  - \undemi \lambdad ikab  \F jc . \tXd abij  
-  \lambdad ikeb   \W ejac . \tXd abij  + \lambdad mjac   \W ikmb . \tXd abij  \nonumber \\
& -\unquart \lambdad ijec   \W ekab . \tXd abij  + \unquart \lambdad mkab   \W ijmc . \tXd  abij  + \undemi \lambdad ijae  \W ekbc . \tXd abij  - \undemi \lambdad imab  \W jkmc . \tXd abij  \\
\displaybreak[1]
&\nonumber\\
\left( \tXs {}{} F \right)^{kl}_{cd} =&+ \Pmm cd \Pmm kl \lambdas kc   \V lida . \tXs ai - \Pmm kl \lambdas ka  \V ilcd . \tXs ai  - \Pmm cd\lambdas ic  \V klad . \tXs ai  -\Pmm kl \lambdad kicd \F la . \tXs ai  - \Pmm cd  \lambdad klca  \F id . \tXs ai \nonumber \\
& + \Pmm cd \lambdad klce  \W eida . \tXs ai  - \Pmm kl \lambdad kmcd  \W lima . \tXs ai 
 + \Pmm cd \Pmm kl \lambdad mlca    \W kimd . \tXs ai  + \lambdad micd  \W klma . \tXs ai  \nonumber \\
 &-  \Pmm cd \Pmm kl \lambdad kied  \W elca . \tXs ai  - \lambdad klea \W eicd . \tXs ai    -\undemi \Pmm cd \lambdad klca  \V ijdb . \tXd abij   -\undemi \Pmm kl\lambdad kicd  \V ljab . \tXd abij  +\unquart \lambdad ijcd  \V klab . \tXd abij \nonumber \\
 & + \unquart \lambdad klab  \V ijcd  . \tXd abij  +\Pmm kl \Pmm cd \lambdad kjcb  \V lida . \tXd abij  -\undemi \Pmm kl \lambdad ikab \V jlcd . \tXd abij -\undemi \Pmm cd \lambdad ijac \V klbd . \tXd abij 
\end{align}

\subsection{EOM-EE $\sigma$-Vectors and intermediates }


The programmable expressions for the elements for the EOM-EE right $^{R}\sigma$ and left $^{L}\sigma$ vectors are given by:
\begin{align}
    ^{R}\sigma_{i}^{a} = &     \F ae r_{i}^{e}    -\F mi r_{m}^{a}+   \F me r_{mi}^{ea}    +\W ei ma  r_{m}^{e}     +\frac{1}{2} \W am ef r_{im}^{ef}
    +\frac{1}{2} \W mn ie r_{mn}^{ea} \label{sigma_sin-ee} \\
    ^{R}\sigma_{ij}^{ab} = &- \Pmm ab  \W mbij \rs am +\Pmm ij \W abej \rs ei + \Pmm ab (\W bmfe \rs em )\td afij -P_{-ij}(\W nmje \rs em )\td abin \nonumber  \\
    &+P_{-ab}\F be \rd aeij - \Pmm ij \F mj \rd abim +\frac{1}{2} \W mnij \rd abmn  + \Pmm ab \Pmm ij \W mbej \rd aeim \nonumber  \\
    &-\Pmm ab \frac{1}{2}(\V nmfe \rd eamn )\td fbij -\Pmm ij \frac{1}{2}(\V nmfe \rd feim )\td bajn +\frac{1}{2} \W abef \rd efij \label{sigma_double-ee}\\
    &\nonumber \\
     ^{L}\sigma_{a}^{i} =& \ls ie \F ea - \ls ma \F im + \frac{1}{2} \ld imef \W efam -\undemi \ld mnae \W iemn  - G_{e}^{f} \W eifa -  G_{m}^{n} \W mina +\ls me \W ieam \label{sigma_sin-ee-left} \\
    ^{L}\sigma_{ab}^{ij}=& \Pmm ab \ld ijae \F eb -\Pmm ij \ld imab \F jm + \undemi \ld mnab \W ijmn + \Pmm ij \Pmm ab \ld imae \W jebm + \Pmm ab \V ijae G_{b}^{e} - \ls ma \W ijmb  \nonumber \\
    & -\Pmm ij \V imab  G_{m}^{j} + \Pmm ij \ls ie \V ejab + \Pmm ij \Pmm ab \ls ia \F jb + \undemi \ld ijef \W efam \label{sigma_double-ee-left}
 \end{align}

\clearpage
The programmable expressions for the elements of the intermediates $\F {}{},\ \W{}{}{}{} $  used above are given by:
\begin{align}
 	\F im = & f_{m}^{i} +f_{e}^{i}t_{m}^{e}+V_{me}^{in}t_{n}^{e}+\frac{1}{2}V_{ef}^{in}\tau_{mn}^{ef} 	\label{F_im} \\
	\F ae = & f_{e}^{a} -f_{a}^{m}t_{m}^{e}+V_{fa}^{me}t_{m}^{f}-\frac{1}{2}V_{af}^{mn}\tau_{mn}^{ef} 	\label{F_ea}\\
	\F me = & f_{e}^{m} + V_{ef}^{mn}t_{n}^{f}	\label{F_me} \\
	\W ijmn = & V_{mn}^{ij} + P_{-mn}V_{en}^{ij}t_{m}^{e}+\frac{1}{2}V_{ef}^{ij}\tau_{mn}^{ef} 	\label{W_ijmn} \\
    \W mbej = & V_{ej}^{mb} + V_{ef}^{mb}t_{j}^{f}-V_{ej}^{mn}t_{n}^{b}-V_{ef}^{mn}(t_{jn}^{fb}+t_{j}^{f}t_{n}^{b})
	\label{W_mbej}\\
	\W iemn = & V_{mn}^{ie} + \F if t_{mn}^{ef}- \W iomn t_{o}^{e}+\frac{1}{2}V_{fg}^{ie}\tau_{mn}^{fg} \nonumber \\
	&+P_{-mn}\bar{W}_{fn}^{ie}t_{m}^{f}+P_{-mn}V_{mf}^{io}t_{no}^{ef} 	\label{W_iemn} \\
	\W efam = &V_{am}^{ef} + P_{-ef}V_{ag}^{en}t_{mn}^{gf} \nonumber\\
	&+\W efag t_{m}^{g}+\F na t_{mn}^{ef}+\frac{1}{2}V_{am}^{no}\tau_{no}^{ef}-P_{-ef}\bar{W}_{am}^{nf}t_{n}^{e}
	\label{W_efam}\\
	\W efab = &V_{ab}^{ef} - P_{-ef}V_{ab}^{mf}t_{m}^{e}+\frac{1}{2}V_{ab}^{mn}\tau_{mn}^{ef}
	\label{W_efab}\\
	\bar{W}_{ej}^{mb} =& V_{ej}^{mb}-V_{ef}^{mn}t_{nj}^{bf}
	\label{barW_mbej}\\
	\W mnie =& V_{ie}^{mn} + t_{i}^{f}V_{fe}^{mn}
    \label{W_mnie}\\
	\W amef = & V_{ef}^{am} - V_{ef}^{nm}t_{n}^{a}
	\label{W_amef}\\
	G_{a}^{e} = & -\frac{1}{2}l_{af}^{mn}t_{mn}^{ef}
	\label{G_ea}\\
	G_{m}^{i} = &\frac{1}{2}l_{ef}^{in}t_{mn}^{ef}
	\label{G_mi}\\
	\tau^{ab}_{ij} =& \td abij + \undemi \Pmm ab \Pmm ij \ts ai \ts bj
	\label{tau}
\end{align}

\clearpage

\section{Davidson scheme for solving first-order response equation}

Due to the extremely large dimension of $\bar{\mathbf{H}}$ in practical calculations, solving the response equations to obtain the perturbed amplitudes calls for the use of iterative procedures~\cite{er1975iterativecalculationof,hirao1982generalization}, since directly inverting $(\bar{\mathbf{H}} - \omega_k\mathbf{I})$
in the full single and double excitation space to is not feasible in all but the simplest cases. 
In this work we have opted to follow the scheme outlined by~\citet{olsen1988solution}, with adjustments due to the fact that in $\bar{\mathbf{H}}$ is non-symmetric, so that a common framework for solving both linear systems and eigenvalue equations can be put in place. 

To this end, we have reimplemented and generalized the Davidson solver code outlined by~\citet{shee2018equation}, so that all matrix/vector operations are now expressed in terms of tensor operations, involving the tensor datatypes available in the GPU-accelerated tensor operation frameworks used in ExaCorr~\cite{pototschnig_implementation_2021}. In doing so, we have conserved the features previously implemented for the solution of eigenvalue equations (multi-root solutions, root-following, etc), and added the ability to solve right-hand side and left-hand side linear systems (though for linear response we will only make use of the righ-hand side solutions).  

\begin{figure}[htb!]
     \centering
     \includegraphics[width=16.5cm,height=12.5cm]{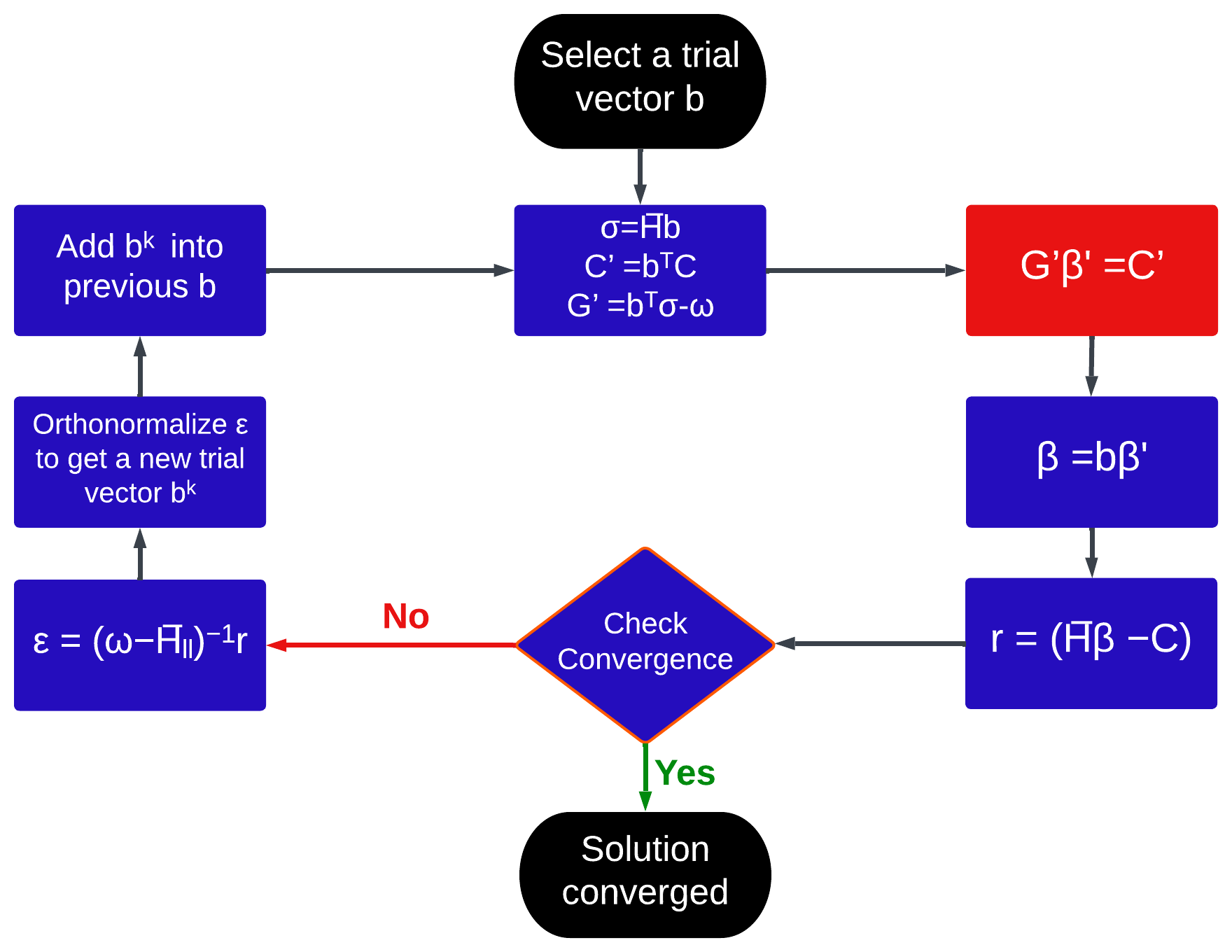}
     \caption{Workflow of the Davidson scheme for solving response equations. The operations performed within the black boxes are independent from the Davidson loop. The tasks in the blue boxes are implemented using TAL-SH tensors. The solution of the linear system in the subspace, indicated by the red box, employs Fortran arrays.  
     }
     \label{fig:davidson}
 \end{figure}

The iterative solver workflow for the solution of linear systems therefore consists of the following steps, which are summarized in figure~\ref{fig:davidson}:
\begin{enumerate}
    \item Choose an orthonormal vector as the initial guess for the trial vector space $\{\mathbf{b}\}$ where $\mathbf{t}=\mathbf{b}\boldsymbol{\beta}^\prime$ (note that $\mathbf{t}$ and $\mathbf{b}$ are TAL-SH tensors and $\boldsymbol{\beta}^\prime$ is a Fortran array). By default, we start with pivoted unit trial vectors (see below for details);
    \item Construct the reduced subspace matrix  $\mathbf{G}^\prime$ and column vector  $\mathbf{C'}$ by projecting the $\mathbf{G} = (\bar{\mathbf{H}} - \omega_k\mathbf{I})$ matrix $\bar{\mathbf{H}}$ and property gradient vector $\mathbf{C} = \boldsymbol{\xi}^{Y}$ onto the current trial vector space $\{\mathbf{b}_i, i = 1,\ldots L\}$, respectively. The $\bar{\mathbf{H}}\mathbf{b}$ products are obtained with the EOM-EE $\boldsymbol{\sigma}$-vector routines;
    \begin{equation}
    \mathbf{G}\mathbf{t}^Y=-\boldsymbol{\xi}^Y \implies \mathbf{b}^\dagger \mathbf{G}\mathbf{b}\boldsymbol{\beta}^\prime=\mathbf{b}^\dagger\mathbf{C} \implies  \mathbf{G}^\prime\boldsymbol{\beta}^\prime = \mathbf{C}^\prime
    \end{equation}
    \item Evaluate the residual vector ($\mathbf{r}^k$) and preconditioner ($\mathbf{p}^k$) 
    \begin{align*}        
    \mathbf{r}^k&= (\boldsymbol{\sigma}- \omega_k\mathbf{b})\boldsymbol{\beta}^\prime - \mathbf{C} \\
    \mathbf{p}^k &= (\omega_k-\bar{H}_{||})^{-1}
    \end{align*}
    the latter being utilized to improve convergence~\cite{saue_linear_2003}; compute the norm of $\mathbf{r}^k$ and compare it to the threshold defined by the user ;
    \item If the norm of $\mathbf{r}^k$ exceeds the threshold, it indicates that the calculation has not converged. In this case, we construct the correction vector $\boldsymbol{\epsilon}^k = \mathbf{p}^k\mathbf{r}^k$ and orthonormalize it to the existing trial vector using a modified Gram-Schmidt procedure, in order to generate the new trial vectors $\mathbf{b}^{k}$, adding it to $\{\mathbf{b}_i, i=1,\ldots,L,L_k\}$. Using the newly generated trial vectors, repeat step 2 until the norm of the residual vector becomes smaller than the threshold;
    \item Once the norm of the error vector falls below the threshold, it indicates that the calculation has converged. At this point, the Davidson routine stops, and the final solution vector is obtained from $\mathbf{b}$ and $\boldsymbol{\beta}^\prime$ as a TAL-SH tensor, which is subsequently used to calculate the response function of interest.
\end{enumerate}

\clearpage

\section{Frequency dependent polarizability of IIB atoms}
\begin{figure}[H]
    \centering
    \includegraphics[width = .48\linewidth,height=6.5cm]{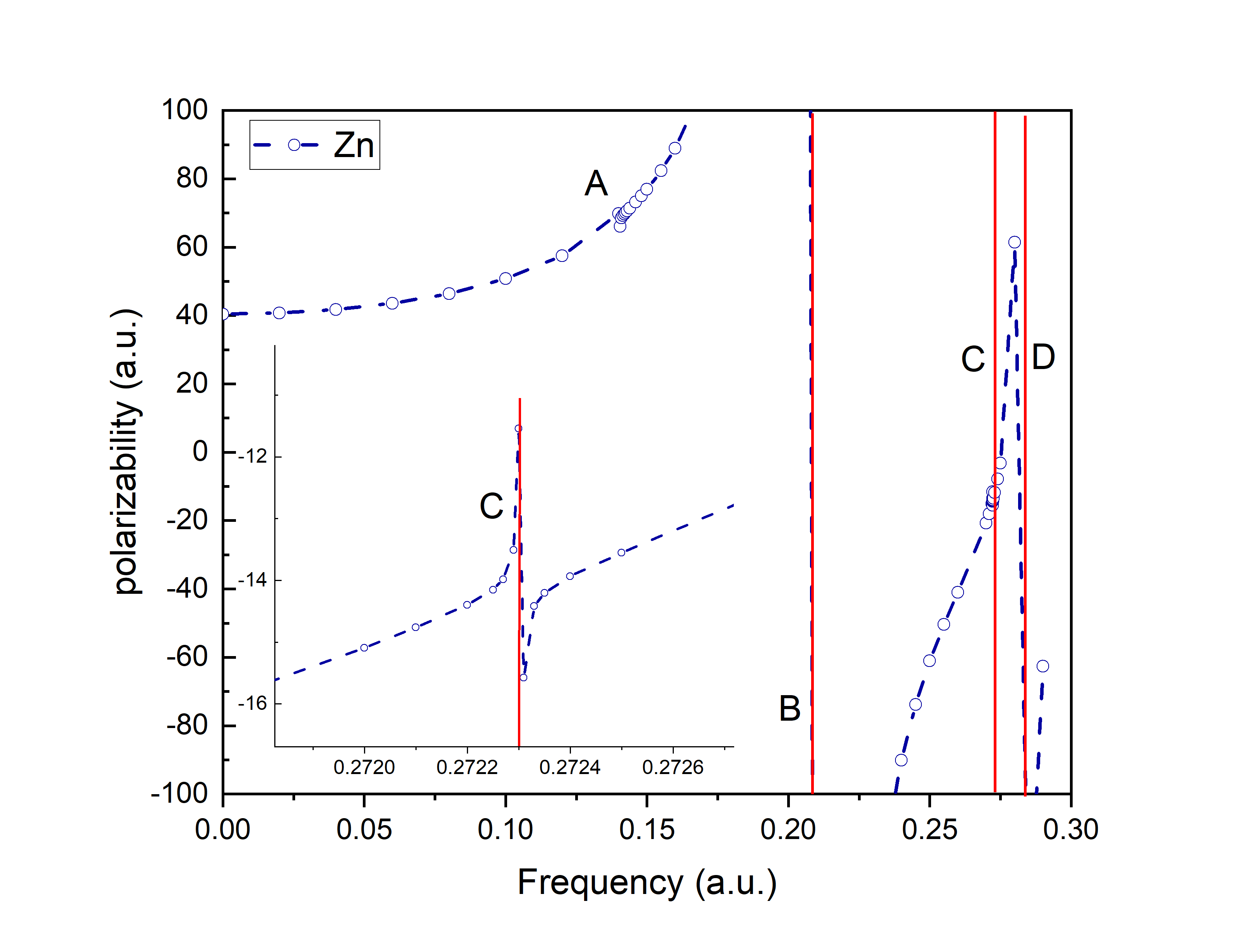}
    \includegraphics[width = .48\linewidth,height=6.5cm]{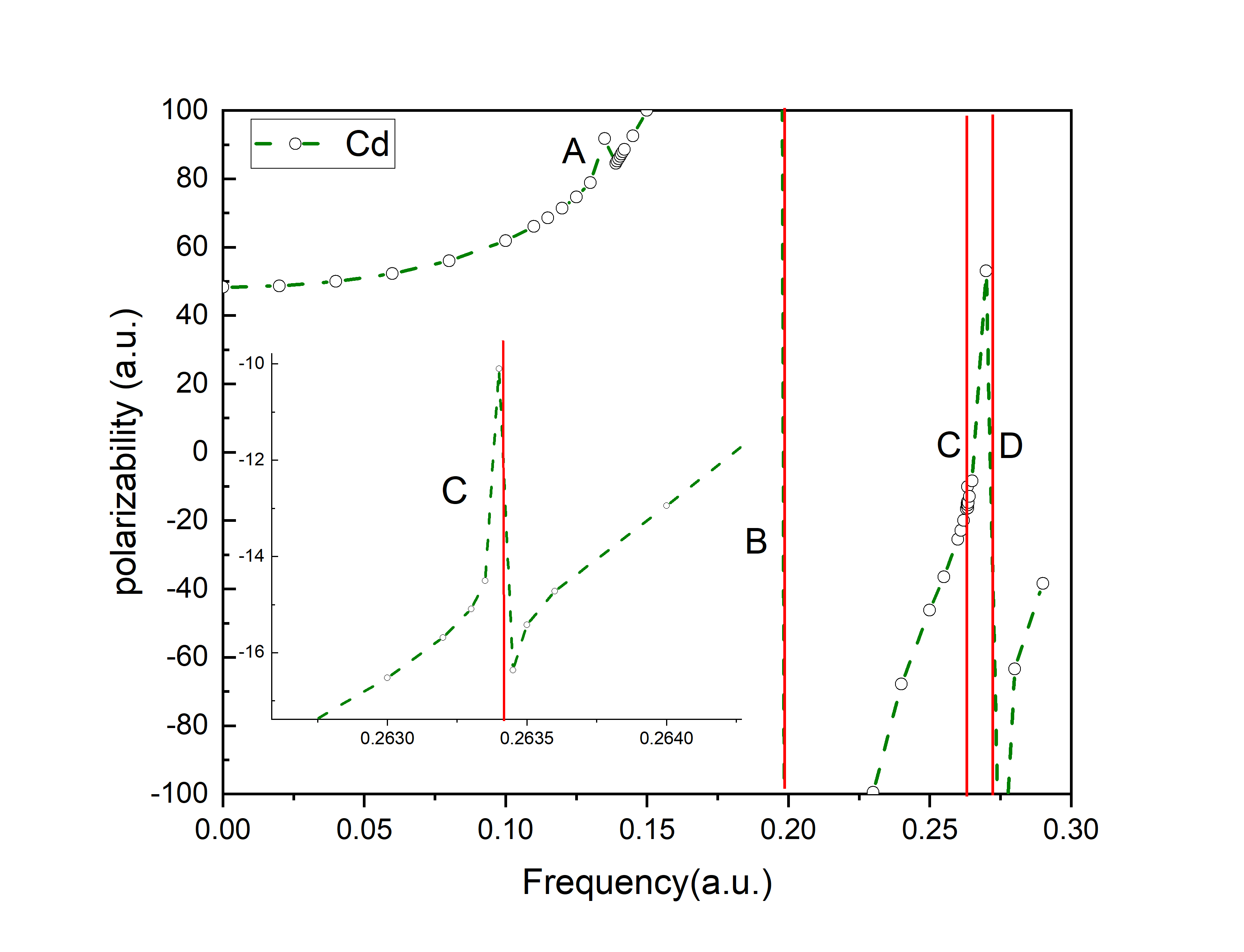}
    \caption{Frequency dependent polarizability of Zn and Cd.}
    \label{fig:fre-dep-iib-c}
\end{figure}

      \begin{table}[H]
        \begin{threeparttable}
            
            \center
            \caption{Number of iterations for solving response equations with different frequencies}\label{tab:iib_iternative}

        \setlength{\tabcolsep}{11.0mm}{
            \begin{tabular}{cccc}
            \hline
            Frequency (a.u.)&  Zn &Cd &Hg \\
            \hline
            0.0 &15 &14 &13 \\
            0.05 &18 &16 &14 \\
            X\tnote{a} &49 &39 &52 \\
             X\tnote{b} &39 &36 &41 \\
            \hline
            \end{tabular}}
            \begin{tablenotes}
                \item[a] Frequency close to the \textbf{B} transition: For \textbf{Zn}: 0.213 a.u.; For \textbf{Cd}: 0.203 a.u; For \textbf{Hg}: 0.25 a.u; 
                \item[b]: With damping factor $\gamma$ = 0.01 a.u.
            \end{tablenotes}

        \end{threeparttable}
        \end{table}

\clearpage

\section{Comparison between LRCC with DIRAC and Dalton}

\begin{figure}
    \centering
    \includegraphics[width=17cm,height=13cm]{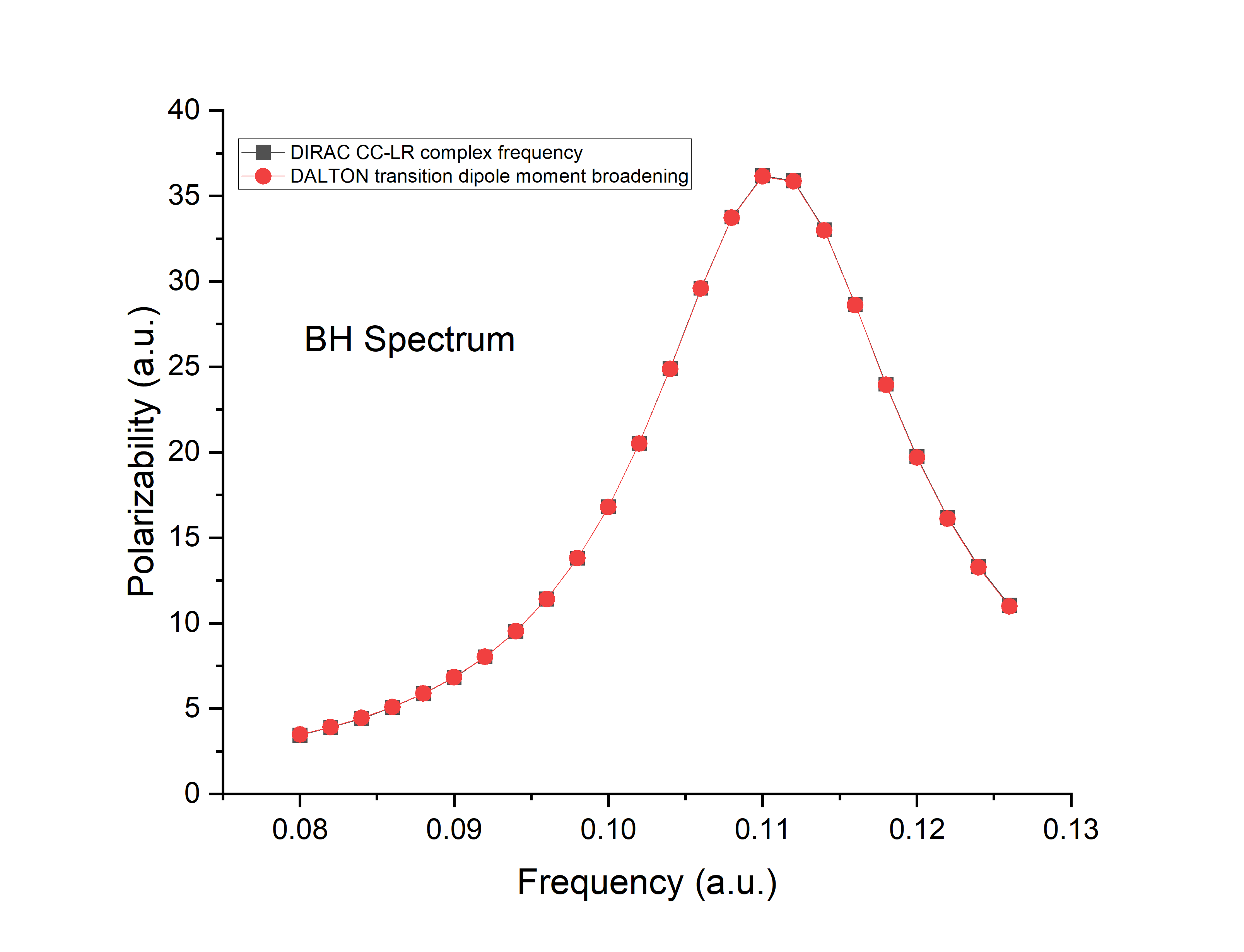}
    \caption{Comparison of LR-CC damped response results obtained with DIRAC with the Levy-Leblond Hamiltonian and standard LR-CC results obtained with Dalton, to which a Lorentzian broadening $ L=\frac{1}{1+(\frac{\omega-\omega_{0}}{\gamma/2})^{2}},$ with the same $\gamma$ used in the LR-CC damped response calculations has been applied.}
    \label{fig:my_label}
\end{figure}
\clearpage

\section{Excitation energy of I$_{2}$}

We present the results of excitation energy of I$_{2}$ for  B$^{3}\Pi_{0+}$ and C$^{1}\Pi_{1}$ states. We use the RELCCSD module to perform the calculations, where all occupied and virtual orbitals are correlated.  The Complete basis set(CBS) results are extrapolated by TZ and QZ values with the formula:
    \begin{equation*}
        E_{CBS}=\frac{4^3E_{QZ}-3^3E_{TZ}}{4^3-3^3}
    \end{equation*}

        \begin{table}[H]
%
            \caption{EOM-CCSD Excitation energy of I$_{2}$ for B$^{3}\Pi_{0+}$ and C$^{1}\Pi_{1}$ states\label{tab:i2}}
        \setlength{\tabcolsep}{10.0mm}{
            \begin{tabular}{ccccc}
            \hline
            &  \multicolumn{2}{c}{B$^{3}\Pi_{0+}$} &\multicolumn{2}{c}{C$^{1}\Pi_{1}$} \\
            \hline
            &  eV & nm & eV &nm\\
            \hline
            aug-DZ &2.28 &544 &2.54 &487 \\
            aug-TZ &2.38 &520 &2.59 &478 \\
            aug-QZ &2.41 &514 &2.61 &475 \\
            CBS &2.43 &510 &2.62 &473 \\
            Exp &2.34 &530 &2.51 &495 \\
            \hline
            \end{tabular}}
        \end{table}

\clearpage

\section{Indirect spin-spin couplings}

We present here the results for the Indirect spin-spin coupling $J$ for the hydrogen halides.

        \begin{table}[H]
        \begin{threeparttable}
            
            \centering
            \caption{Isotropic and anisotropic spin-spin coupling J(Hz) for HX(X=F, Cl, Br, I)\label{tab:spsp}}

        \setlength{\tabcolsep}{8.0mm}{
            \begin{tabular}{ccccc}
            Models&  $^{1}$HF$^{19}$ &$^{1}$HCl$^{35}$ &$^{1}$HBr$^{79}$ &$^{1}$HI$^{127}$ \\
            \hline
             \multicolumn{5}{c}{Isotropic}\\
            \hline
            NR-HF     &560.2324 &33.1722 &32.6909 &-2.1729\\
            NR-B3LYP  &377.5294 &23.2295 &-5.6679 &\\
            NR-CC-CI  &458.5500 &36.9018 &92.9897 &\\
            NR-CC-CC  &457.6892 &36.6414 &90.5146 &\\
            NR-CC-CC\tnote{a} &457.6715 &36.6414 &90.5144 &\\
            X2C-HF    &559.7091 &32.0803 &-13.6915 &-201.2263\\
            X2C-B3LYP &375.7986 &22.3179 &-35.3065 &-138.7448\\
            X2C-CC-CI &457.7350 &36.4102 &71.9471 &8.4426\\
            X2C-CC-CC &456.8450 &36.1321 &68.7287 &1.8104\\
            DC-HF     &559.3715 &31.9899 &-14.6151 &-203.2970\\
            \hline
            \multicolumn{5}{c}{Anisotropic}\\
            \hline
            NR-HF     &28.8305 &70.3046 &489.1606 &672.0842\\
            NR-B3LYP  &71.7803 &59.0412 &393.8655 &\\
            NR-CC-CI  &-42.4747 &42.8696 &304.9419 &\\
            NR-CC-CC  &-39.4953 &43.7373 &310.8541 &\\
            X2C-HF    &29.2373 &70.9774 &508.9765 &739.8355\\
            X2C-B3LYP &72.9026 &59.3847 &394.2909 &486.5593\\
            X2C-CC-CI &-41.3592 &43.4599 &322.0869 &466.6810\\
            X2C-CC-CC &-38.3627 &44.3369 &328.3281 &475.1557\\
            DC-HF     &29.3720 &71.0300 &509.5169 &740.8219\\
            \hline           
            \end{tabular}}
            \begin{tablenotes}
              \item[a] Calculations were performed using the CFOUR program
            \end{tablenotes}
        \end{threeparttable}
        \end{table}

        \clearpage

      \begin{table}[H]
        \begin{threeparttable}
          \center
            \caption{Difference in the linear response contribution to the spin-spin coupling constant $K_{HX}$ between CC-CI and CC-CC ($\Delta K^\text{LR} = K^\text{LR}(\text{CC-CI}) - K^\text{LR}(\text{CC-CC})$, in a.\ u.) for the HX(X=F, Cl, Br, I) systems, broken down into the $xx, yy, zz$ components. Apart from the absolute values, we also provide the difference per correlated electron ($\Delta K^\text{LR}_e = \Delta K^\text{LR} / N_e$, with $N_e = 10, 18, 36$ and $54$ across the series)}\label{tab:spsp-ccci-vs-cccc}

        \setlength{\tabcolsep}{6.0mm}{
            \begin{tabular}{cccccc}
                  & &             &\multicolumn{3}{c}{components} \\
\cline{4-6}
         &   System & Hamiltonian & ${zz}$ & ${xx}$  & ${yy}$ \\
            \hline
$\Delta K^\text{LR}$
&	HF	&	NR &	-0.0277	&	0.0456	&	0.0456	\\
&		&	X2C	&	-0.0273	&	0.0465	&	0.0465	\\
&	HCl	&	NR	&	-0.0751	&	0.1298	&	0.1298	\\
&		&	X2C	&	-0.0724	&	0.1347	&	0.1347	\\
&	HBr	&	NR	&	-0.1351	&	0.4097	&	0.4097	\\
&		&	X2C	&	-0.0860	&	0.4834	&	0.4834	\\
&	HI	&	X2C	&	0.1130	&	1.0875	&	1.0875	\\            \hline
$\Delta K^\text{LR}_e$ 
&	HF	&	NR	&	-0.0028	&	0.0046	&	0.0046	\\
&		&	X2C	&	-0.0027	&	0.0046	&	0.0046	\\
&	HCl	&	NR	&	-0.0042	&	0.0072	&	0.0072	\\
&		&	X2C	&	-0.0040	&	0.0075	&	0.0075	\\
&	HBr	&	NR	&	-0.0038	&	0.0114	&	0.0114	\\
&		&	X2C	&	-0.0024	&	0.0134	&	0.0134	\\
&	HI	&	X2C	&	0.0021	&	0.0201	&	0.0201	\\
\hline           
            \end{tabular}}

        \end{threeparttable}
    \end{table}

    \begin{table}[H]
    \begin{threeparttable}

        \centering
        \caption{Isotropic and anisotropic indirect spin-spin coupling ($J_{iso}$ and $J_{aniso}$ in Hz) for the isolated H$_{2}$Se subsystem taken at the geometry of the supermolecule, ($J_{iso}^{super}$ and $J_{aniso}^{super}$ in Hz) for the H$_{2}$Se subsystem in H$_{2}$Se-H$_{2}$O, and the shifts ($\Delta J$, in Hz) for the isolated ("ME") H$_{2}$Se molecules in the presence of H$_{2}$O}
        \setlength{\tabcolsep}{5.0mm}{
        \begin{tabular}{ccccccc}
        Models  & $J_{iso}^{}$ & $J_{iso}^{super}$ & $\Delta J_{iso}^{ME}$  & $J_{aniso}^{}$ &$J_{aniso}^{super}$ &$\Delta J_{ianso}^{ME}$ \\
        \hline
            \multicolumn{7}{c}{$^{1}$H$_{b}$-H$^{1}$}\\
        \hline
        HF\tnote{a}
             &-17.1157 &-18.4466 &-1.3309 &43.4044 &42.9922  &-0.4122 \\
        HF   &-16.3598 &-17.6472 &-1.2874 &43.5591 &43.1654  &-0.3937 \\
        BLYP & -7.4945 & -7.8597 &-0.3652 &41.2526 &40.9022  &-0.3504 \\
        B3LYP& -8.1176 & -8.5639 &-0.4463 &41.2997 &40.9395  &-0.3602 \\
        CC-CI& -9.8335 &-10.2246 &-0.3911 &39.9376 &39.5455  &-0.3921 \\
        CC-CC&-10.2487 &-10.6565 &-0.4078 &40.1031 &39.7072  &-0.3959 \\
        \hline
            \multicolumn{7}{c}{$^{1}$H-Se$^{34}$}\\
        \hline
        HF\tnote{a}
             & 93.6021 & 93.6042 &0.0021 &302.0356 &300.7717  &-1.2639 \\
        HF   & 55.2216 & 56.1003 &0.8787 &349.7842 &346.5116  &-3.2726 \\
        BLYP &-24.8033 &-21.2742 &3.5291 &263.5003 &261.1415  &-2.3588 \\
        B3LYP& -7.0856 & -4.0854 &3.002  &267.6406 &265.3991  &-2.2415 \\
        CC-CI& 68.1164 & 70.0653 &1.9489 &213.7819 &212.0822  &-1.6998 \\
        CC-CC& 67.4464 & 69.4103 &1.9639 &218.0737 &216.3287  &-1.7450 \\
        \hline
        \end{tabular}}

        \begin{tablenotes}
            \item[a]  Nonrelativistic calculation with the Levy-Leblond Hamiltonian
        \end{tablenotes}
        \label{tab:solvent of h2se}
        \end{threeparttable}
    \end{table}

        \clearpage

\section{Optical rotation}

\begin{table}[H]
            \centering
            \caption{Optical rotation Test of H$_{2}$S$_{2}$ using uncontracted basis set}\label{tab:H2S2opt}

        \setlength{\tabcolsep}{7.5mm}{
            \begin{tabular}{ccc}
            Method    & G Tensor & Optical rotation\\
            \hline
            CC(X2C) & -0.10732 &-271.1273\\
            CC(X2C, virtual to 100 a.u.) & -0.10757&-271.7746 \\
            CC(LEVY-LEBLOND) &-0.10029&-253.3778\\
            CC(DALTON) &-0.10029&-253.3856\\
            \hline
            \end{tabular}}
\end{table}

        \begin{table}[H]
            \centering
            \caption{Optical rotation (a.u.) of Hydrogen peroxide series (H$_{2}$Y$_{2}$) with a frequency corresponding to the sodium D-line (589.29 nm, 0.077319 a.u.) calculated with X2C and LEVY-LEBLOND Hamiltonian}\label{tab:opt-rot-h2y2}

        \setlength{\tabcolsep}{5.5mm}{
            \begin{tabular}{ccccc}
            Method    & H$_{2}$O$_{2}$ & H$_{2}$S$_{2}$& H$_{2}$Se$_{2}$ & H$_{2}$Te$_{2}$\\
            \hline
            HF(LEVY-LEBLOND)    &-93.2588  &-124.5926 &-205.8141  &-93.6808 \\
            HF(X2C)             &-93.0240  &-136.7733 &-22.3730   &2007.9020 \\
            B3LYP(LEVY-LEBLOND) &-172.7622 &-295.4117 &-1098.4716 &-18269 \\
            B3LYP(X2C)          &-173.8253 &-320.9461 &-1418.7842 &4075 \\
            CC(LEVY-LEBLOND)    &-181.5585 &-253.3779 &-386.6950  &-263.5968 \\
            CC(X2C)             &-182.4976 &-271.1273 &-1906.1708 &218.8335 \\
            \hline
            \end{tabular}}
        \end{table}

        \begin{table}[H]
            \centering
            \caption{Excitation energy (a.u.) of the first ten states of H$_{2}$Se$_{2}$ and H$_{2}$Te$_{2}$ from nonrelativistic calculations}\label{tab:h2se2_exc}

        \setlength{\tabcolsep}{8.5mm}{
            \begin{tabular}{ccccc}
            State    & TDHF & CIS & B3LYP & CC\\
            \hline
            \multicolumn{5}{c}{H$_{2}$Se$_{2}$}     \\
            \hline
            \multicolumn{5}{c}{Singlet state}     \\
            \hline
            1    &0.1083 &0.1120 &0.0889  &0.1016 \\
            2    &0.1624 &0.1646 &0.1343  &0.1478 \\
            3    &0.2024 &0.2035 &0.1668  &0.1796 \\
            4    &0.2214 &0.2233 &0.1919  &0.2061 \\
            5    &0.2407 &0.2489 &0.2077  &0.2283 \\
            \hline
            \multicolumn{5}{c}{Triplet state}     \\
            \hline
            1    &0.0719 &0.0820 &0.0661  &0.0812 \\
            2    &0.0954 &0.1311 &0.1121  &0.1309 \\
            3    &0.1315 &0.1461 &0.1510  &0.1561 \\
            4    &0.1798 &0.1841 &0.1576  &0.1703 \\
            5    &0.1900 &0.1958 &0.1764  &0.1901 \\
            \hline
            \multicolumn{5}{c}{Singlet-Triplet Splitting}     \\
            \hline
            1    &0.0364 &0.0300 &0.0228  &0.0203 \\
            2    &0.0671 &0.0335 &0.0222  &0.0169 \\
            3    &0.0709 &0.0574 &0.0158  &0.0235 \\
            4    &0.0416 &0.0392 &0.0343  &0.0359 \\
            5    &0.0507 &0.0531 &0.0313  &0.0383 \\
            \hline
             \multicolumn{5}{c}{H$_{2}$Te$_{2}$}     \\
            \hline
            \multicolumn{5}{c}{Singlet state}     \\
            \hline
            1    &0.0957 &0.0989 &0.0777  &0.0898 \\
            2    &0.1402 &0.1421 &0.1158  &0.1292 \\
            3    &0.1814 &0.1829 &0.1497  &0.1611 \\
            4    &0.1898 &0.1923 &0.1638  &0.1760 \\
            5    &0.1994 &0.2171 &0.1816  &0.1999 \\
            \hline
            \multicolumn{5}{c}{Triplet state}     \\
            \hline
            1    &0.0635 &0.0725 &0.0584  &0.0721 \\
            2    &0.0924 &0.1141 &0.0968  &0.1143 \\
            3    &0.1138 &0.1296 &0.1320  &0.1367 \\
            4    &0.1530 &0.1590 &0.1388  &0.1498 \\
            5    &0.1532 &0.1599 &0.1453  &0.1565 \\
            \hline
            \multicolumn{5}{c}{Singlet-Triplet Splitting}     \\
            \hline
            1    &0.0322 &0.0264 &0.0193  &0.0177 \\
            2    &0.0478 &0.0279 &0.0190  &0.0149 \\
            3    &0.0676 &0.0532 &0.0178  &0.0244 \\
            4    &0.0368 &0.0333 &0.0250  &0.0262 \\
            5    &0.0462 &0.0572 &0.0363  &0.0434 \\
            \hline
            \end{tabular}}
        \end{table}

\clearpage

\section{CPU vs GPU Benchmarks}

We provide below an assessment of the difference in performance between RelCCSD, CPU and GPU-based execution of EOM-EE and response equation of Hg atoms with the same correlation orbital space and basis set indicated in the manuscript: 12 occupied and 102 virtual spinors. 

These calculations have all been carried out in a single node of the TGCC BULL Irene/Joliot Curie (partition V100L), consisting of a Intel(R) Xeon(R) Gold 6240 CPU @ 2.60GHz (72 cores and 386 Gb of RAM per node) and one NVIDIA Tesla V100-PCIE-32GB GPU. To make the comparision only dependent upon the GPU usage, we have set the number of OpenMP threads to 1 in all runs, and toggle between CPU and GPU usage for the same executable. Both ExaCorr and the standalone code were compiled with GCC 11.1.0, CUDA 11.7 and Openblas 0.3.15 as available in the computer center.

To examine this in greater detail, we focus on the single evaluation of the EOM-EE sigma vector, which is central to the procedure of solving the response equations (as well as in the determination of EOM eigenvalues and eigenvectors). We have done so with a stand-alone program that sets up, for a given problem size (number of occupied and virtual spinors), the necessary variables (1-/2-body integral tensors, EOM-CC intermediates, T amplitudes and a single trial vector) and executes a single instance of the production EOM-EE sigma vector code 10 times over, while collecting at each evaluation the elapsed walltime. The results for these evaluations is presented in table~\ref{tab:sigma-benchmark}. As is the case with ExaCorr, the stand-alone program does not carry out I/O to disk, with all aforementioned tensors being always present in RAM.

\begin{table}
\caption{Timings (in seconds) for benchmark evaluations of the EOM-EE sigma vector code used by the response equation solver using CPU and GPU (and the speedup for the GPU case) for different problem sizes (number of occupied \textbf{O} and virtual \textbf{V} spinors), as well as a model metric for the cost \textbf{C} of the evaluation (calculated as the formal O$^2$V$^4$ scaling). For convenience, a scaled cost (\textbf{SC}, obtained by scaling all \textbf{C} values by the smallest one) is also provided.\label{tab:sigma-benchmark}}
\begin{tabular}{rrrrrrr}
\hline
\hline
            &                & \multicolumn{2}{c}{time (s)} && \multicolumn{2}{c}{Cost metric} \\
            \cline{3-4}\cline{6-7}
\textbf{O}   &   \textbf{V}   &  CPU  &  GPU  & speedup &  \textbf{C} & \textbf{SC}	\\								
\hline
10	&	102	&	6	&	1	&	6	&	1.08E+10	&	1.0	\\
10	&	130	&	13	&	2	&	6	&	2.86E+10	&	2.6	\\
10	&	160	&	26	&	5	&	6	&	6.55E+10	&	6.1	\\
10	&	200	&	58	&	10	&	6	&	1.60E+11	&	14.8	\\
\hline
20	&	102	&	20	&	2	&	9	&	4.33E+10	&	4.0	\\
20	&	130	&	43	&	4	&	10	&	1.14E+11	&	10.6	\\
20	&	160	&	85	&	8	&	10	&	2.62E+11	&	24.2	\\
20	&	200	&	183	&	16	&	12	&	6.40E+11	&	59.1	\\
\hline
30	&	102	&	48	&	5	&	10	&	9.74E+10	&	9.0	\\
30	&	130	&	103	&	8	&	13	&	2.57E+11	&	23.7	\\
30	&	160	&	203	&	14	&	15	&	5.90E+11	&	54.5	\\
30	&	200	&	429	&	25	&	17	&	1.44E+12	&	133.0	\\
\hline
40	&	102	&	96	&	8	&	12	&	1.73E+11	&	16.0	\\
40	&	130	&	203	&	13	&	15	&	4.57E+11	&	42.2	\\
40	&	160	&	395	&	21	&	19	&	1.05E+12	&	96.9	\\
40	&	200	&	821	&	37	&	22	&	2.56E+12	&	236.5	\\
\hline\hline
\end{tabular}
\end{table}

From these results, we observe speedups of at least a factor of 6 for smaller problem sizes (which are comparable to the setup for the valence-only calculation on the Hg atom described in the paper), though in absolute terms sigma vector evaluation is rather quick for both CPU and GPU execution. For larger problem sizes, we observe somewhat larger speedups for the GPU code.

\begin{figure}[H]
    \centering
    \includegraphics[width = .75\linewidth,height=8.0cm]{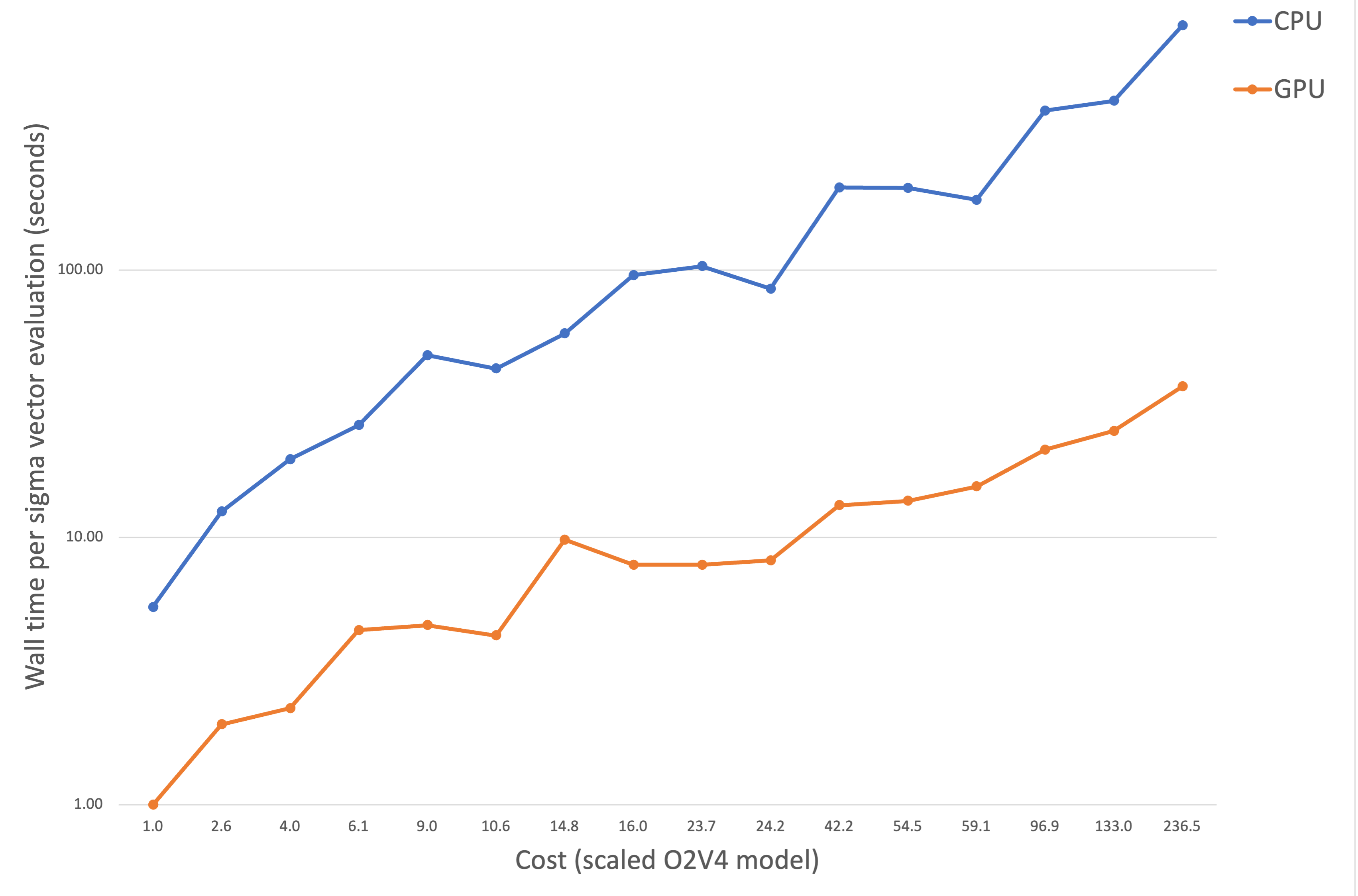}
    \caption{Scaling of sigma vector evaluation for benchmark calculations employing CPU and GPU offloading for increased problem size (on the basis of the scaled cost measure, see table~\ref{tab:sigma-benchmark}).}
    \label{fig:sigma-benchmark}
\end{figure}

The difference in performance reported on table~\ref{tab:sigma-benchmark} can also be visualized in terms of the scaling plot, shown in figure~\ref{fig:sigma-benchmark}, which presents the increase execution time with respect to the increase in problem size, here taken to be the \textbf{SC} cost metric presented above. From the figure we see for both CPU and GPU code execution time grows in a fairly linear manner (in log scale) with increased problem size. 

We note the slope for the CPU case is somewhat larger than for the GPU code, which could indicate the increase in speedup with problem size could come at least in part from a less optimized CPU implementation as opposed to being purely a gain in performance of the GPU code with increased problem size, but at this stage we have not explored the matter further. We were unable to increase the problem size beyond those shown due to the amount of RAM available on the node.

Beyond assessing the performance of the code for sigma vector evaluation, we present in table~\ref{tab:hg-benchmark} a comparison of the CPU and GPU performance for the solution of the linear response functions to determine the XX, YY and ZZ components of the dipole polarizability for the mercury atom. This comparison provides a picture closer to the actual code use in production. 

Our results indicate that GPU offloading remains advantageous in practice as a means to speed up our calculations. However, we see that for the tests carried out the speedup in the total time to solution in each of these code sections is not as significant as for the sigma vector evaluation itself. This is not unexpected since there are other operations which are less computationally intensive than the sigma vector evaluation (such as trial vector update, orthonormalization and antisymmetrization). That said, as the problem size increased we see an increase of the speedup, in line with what is observed for the sigma vector evaluation.

\begin{table}
\caption{Walltimes (in seconds) for the aggregate time for determination of the XX, YY and ZZ components of the dipole polarizability (LR) for the Hg atom for GPU and CPU execution, as a function of  problem size (number of occupied \textbf{O} and virtual \textbf{V} spinors). For a direct comparison to the results in table~\ref{tab:sigma-benchmark}, all calculations employed 1 OpenMP thread.\label{tab:hg-benchmark}}
\begin{tabular}{rrrrr}
\hline
\hline
            &      & \multicolumn{2}{c}{time (s)} \\
            \cline{3-4}
\textbf{O} & \textbf{V} &  CPU  &  GPU  & speedup \\								
\hline
12	&	102	&	773 &  317 & 2.4 \\
18  &   148 &  5485 & 1524 & 3.6\\
18  &   180 &10879 &2606 &4.2\\
\hline\hline
\end{tabular}
\end{table}

\section{ExaCorr vs RELCCSD comparison}

We present below a comparison between the ExaCorr and RELCCSD implementations. As response theory is not implemented in RELCCSD, the fairest comparison between codes that can be done is for the solution of the EOM-EE problem for a single excited state, and without making use of double group point symmetry in RELCCSD. Given the very different design decisions (basically a tradeoff between storing data on RAM for ExaCorr and on disk for RELCCSD), we compare only the time to solution for the EOM-EE equations in each case. Our results, shown in table \ref{tab:relccsd-exacorr}, indicate that time to solution between RELCCSD and ExaCorr (GPU) are comparable, though for the small problem size considered here RELCCSD still shows a smaller time to solution. 

\begin{table}[H]
    \caption{Walltimes (in seconds) for the aggregate time for determination of a single EOM-EE excited state for Hg electron for the Hg atom, employing the RELCCSD and ExaCorr (CPU and GPU) codes, for a given number of occupied \textbf{O} and virtual \textbf{V} spinors.\label{tab:relccsd-exacorr}}
    \begin{tabular}{ccccc}
    \hline
\textbf{O} & \textbf{V} & RELCCSD & ExaCorr (CPU) & Exacorr(GPU) \\
     \hline
12 & 102  & 46  &255  &115  \\
    \hline
    \end{tabular}
\end{table}

\bibliography{ms}